\documentclass[sigconf]{acmart} 

\AtBeginDocument{%
  }

\copyrightyear{2024}
\acmYear{2024}
\setcopyright{acmlicensed}\acmConference[CHI '24]{Proceedings of the CHI Conference on Human Factors in Computing Systems}{May 11--16, 2024}{Honolulu, HI, USA}
\acmBooktitle{Proceedings of the CHI Conference on Human Factors in Computing Systems (CHI '24), May 11--16, 2024, Honolulu, HI, USA}
\acmDOI{10.1145/3613904.3642868}
\acmISBN{979-8-4007-0330-0/24/05}

\usepackage{color}
\usepackage{graphicx}
\usepackage{subfigure} 
\usepackage{booktabs} 
\usepackage{multirow}
\usepackage{placeins}
\usepackage{float}

\begin{document}

\title{ReelFramer: Human-AI Co-Creation for News-to-Video Translation}

\author{Sitong Wang}
\affiliation{
  \institution{Columbia University}
  \city{New York}
  \state{NY}
  \country{USA}
}
\email{sw3504@columbia.edu}

\author{Samia Menon}
\affiliation{
  \institution{Columbia University}
  \city{New York}
  \state{NY}
  \country{USA}
}
\email{sm4788@columbia.edu}

\author{Tao Long}
\affiliation{
  \institution{Columbia University}
  \city{New York}
  \state{NY}
  \country{USA}
}
\email{long@cs.columbia.edu}

\author{Keren Henderson}
\affiliation{
  \institution{Syracuse University}
  \city{Syracuse}
  \state{NY}
  \country{USA}
}
\email{khenders@syr.edu}

\author{Dingzeyu Li}
\affiliation{
  \institution{Adobe Research}
  \city{Seattle}
  \state{WA}
  \country{USA}
}
\email{dinli@adobe.com}

\author{Kevin Crowston}
\affiliation{
  \institution{Syracuse University}
  \city{Syracuse}
  \state{NY}
  \country{USA}
}
\email{crowston@g.syr.edu}

\author{Mark Hansen}
\affiliation{
  \institution{Columbia University}
  \city{New York}
  \state{NY}
  \country{USA}
}
\email{mh3287@columbia.edu}

\author{Jeffrey V. Nickerson}
\affiliation{
  \institution{Stevens Institute of Technology}
  \city{Hoboken}
  \state{NJ}
  \country{USA}
}
\email{jnickers@stevens.edu}

\author{Lydia B. Chilton}
\affiliation{
  \institution{Columbia University}
  \city{New York}
  \state{NY}
  \country{USA}
}
\email{chilton@cs.columbia.edu}

\renewcommand{\shortauthors}{Wang, et al.}

\begin{abstract}
Short videos on social media are the dominant way young people consume content.
News outlets aim to reach audiences through news reels---short videos conveying news---but struggle to translate traditional journalistic formats into short, entertaining videos. 
To translate news into social media reels, we support journalists in reframing the narrative. 
In literature, narrative framing is a high-level structure that shapes the overall presentation of a story.
We identified three narrative framings for reels that adapt social media norms but preserve news value, each with a different balance of information and entertainment.
We introduce ReelFramer, a human-AI co-creative system that helps journalists translate print articles into scripts and storyboards.
ReelFramer supports exploring multiple narrative framings to find one appropriate to the story. 
AI suggests foundational narrative details, including characters, plot, setting, and key information.
ReelFramer also supports visual framing; AI suggests character and visual detail designs before generating a full storyboard.
Our studies show that narrative framing introduces the necessary diversity to translate various articles into reels, and establishing foundational details helps generate scripts that are more relevant and coherent.
We also discuss the benefits of using narrative framing and foundational details in content retargeting.
\end{abstract}



\begin{teaserfigure}
\centering
\includegraphics[width=0.94\textwidth]{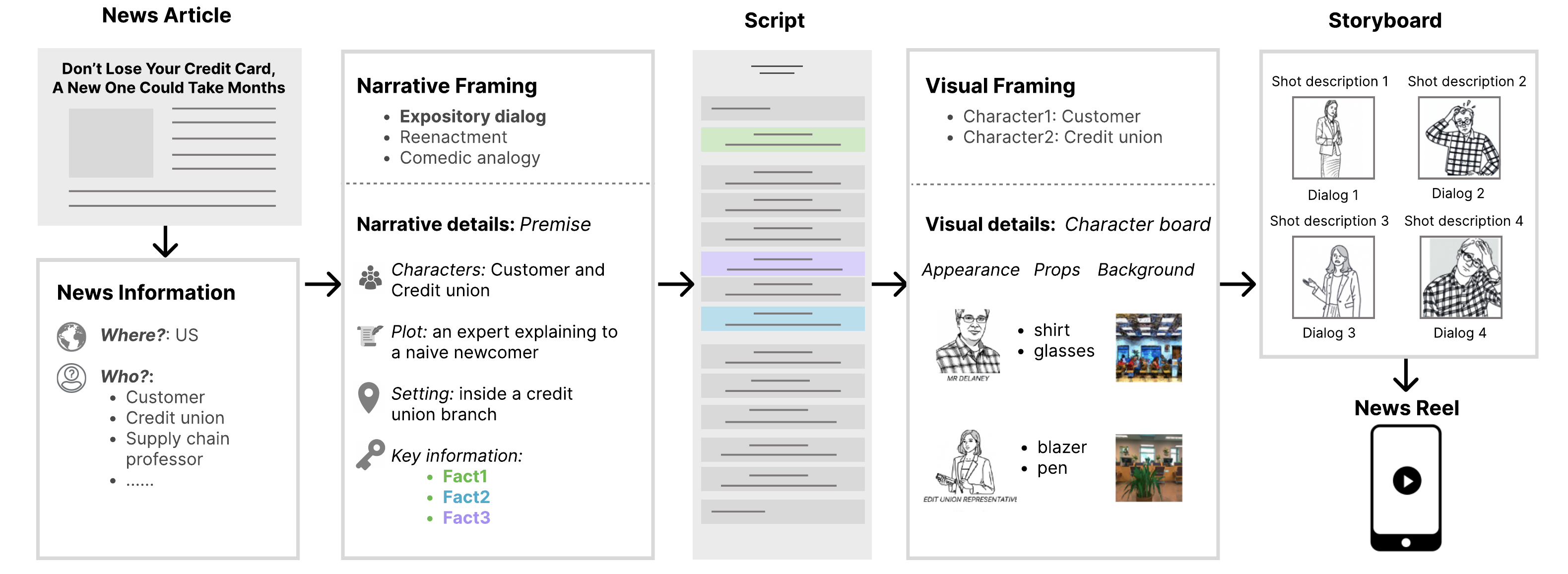}
\caption{
ReelFramer is a human-AI co-creative system that supports journalists in creating news reels by transforming print articles into scripts and storyboards.
ReelFramer enables the exploration of multiple narrative framings that span the infotainment spectrum and their foundational details. 
It also aids in visual framing for character and key visual detail design.
}
\Description{A workflow diagram of ReelFramer, a human-AI co-creative system designed to assist journalists in transforming print articles into news reels. The process begins with the user inputting a news article. Next, the system extracts key information from the article, including the news location, characters, and main events. The user then chooses one of the three narrative framings offered by the system: expository dialog, reenactment, or comedic analogy. Based on the selected framing, the system generates narrative details, including characters, plot, setting, and key information points. Following this, the system creates a script incorporating these narrative details. Once the user is satisfied with the script, the next step involves exploring visual framing for characters. The system generates a character board that provides visual details on the appearance, props, and background of the characters. Building upon the character board, the system creates a storyboard that outlines the shots for the user. This storyboard, alongside the script, serves as a reference for the user to shoot their news reels.}
\label{fig:teaser}
\end{teaserfigure}

\maketitle

\section{Introduction}
Short videos on social media, also known as reels, are quickly becoming the dominant way young people consume information. 
Although platforms like TikTok and Instagram Reels were once used purely for entertainment, such as viral dances, they have also expanded to feature more informational content, including science experiments~\cite{hayes2020making}, how-to videos~\cite{fiallos2021tiktok}, financial literacy~\cite{myrichbff} and news~\cite{newman_digital_news_2022}. 
With the goal of attracting a younger audience, some news outlets are open to using \textit{news reels} as a medium to inform the public. 
\textit{The Washington Post} was the first notable example, and their news reels regularly get over 100,000 views\footnote{\url{https://www.tiktok.com/@washingtonpost}}.
Their fast-paced, story-driven delivery is entertaining as well as informative.
Thus, news reels have the potential to inform young audiences in a format they already know and enjoy. 

News outlets would like to reach audiences through news reels, but currently struggle to translate traditional journalistic formats into short, entertaining videos. 
Traditional journalism typically conveys facts and events with a serious tone and delivers information in a direct manner~\cite{pyramid}.
By contrast, news reels often act out the story from the perspective of different characters that personify aspects of the story. 
For example, in a Washington Post reel\footnote{\url{https://www.tiktok.com/@washingtonpost/video/7147811605655473450}}, one character representing Hurricane Ian explains to another character representing the hurricane’s roommate why he is ``visiting'' Florida and all the trouble he might cause there. 
Even for a serious topic like Hurricane destruction, news reels must strike a balance between informational content and entertainment value~\cite{newman_news_tt_reuters_oxford}.
Exploring different tones and presentations to find the right one for a story can be time-consuming.
Generative AI shows promise in assisting with this exploration but comes with challenges such as maintaining coherence, ensuring the factual accuracy of the generated results~\cite{gpt} and understanding the social media audience. 
AI can potentially be used to aid journalists while mitigating its shortcomings, but journalists still need tools to help them achieve a tone appropriate for social media, yet content worthy of news. 

In literature, narrative framing is a high-level structure that guides the overall presentation of a story and shapes the comprehension of the represented events~\cite{schmid2021narrative}.  
The same story can be told as a drama or a comedy, depending on the framing, choices of characters, and how they interact. 
This concept can also be applied to journalism: the same news story can be presented seriously, as on \textit{The PBS News Hour}, or with comedy, as on \textit{The Daily Show}. 
The choice of framing varies according to the nature of the story; a serious frame might suit some stories, whereas a more light-hearted frame could benefit others. 
Journalists already engage in employing narrative framings as part of the traditional process~\cite{hicks2016writing}. 
However, it is a challenge to develop narrative framings that span the infotainment spectrum and are appropriate for social media reels.

Through a six-month co-design activity with journalism experts, we identified three character-driven narrative framings for reels, which represent different mixtures of information and entertainment:
\textit{Expository dialog} contains the highest informational value, involving an expert explaining information to a newcomer;
\textit{Reenactment} blends information with entertainment, where the main stakeholders act out the event in real time and explore the consequences;
\textit{Comedic analogy} holds the greatest entertainment value, as the news characters perform the event in a comedic setting analogously related to the actual event.
However, a narrative framing can only be executed if the details of the original article fit the details of the framing. 
Thus, we build tools to help users explore different narrative framings and their foundational 
details, including characters, plot, setting, and key information.

We present a human-AI co-creation system called ReelFramer that uses generative text and image AI to scaffold the process of creating a script and storyboard for a news reel (see Figure \ref{fig:teaser}). 
Generative AI is helpful for exploring multiple narrative framings and their details, and generating scripts, character boards, and storyboards.
The system is designed to be highly interactive, allowing people to make choices and guide the system at every step to ensure the content it produces is accurate, authentic, and achievable to execute.
The user starts by inputting the news article.
The system uses a large language model (LLM), GPT-4~\cite{gpt4}, to extract locations, people, and activities from the article. 
The user then selects one of the three narrative framings. 
Given a narrative framing, the system uses the LLM to suggest foundational details like characters, plot, setting, and key information, which is also called \textit{premise} in scriptwriting~\cite{batty2017script}.
The user can accept, regenerate, or edit these suggestions. 
Once the user finalizes a premise, the system automatically generates a script. 
Again, the user can accept, regenerate, or edit the script. 
Users are encouraged to explore many narrative framings, premises, and scripts.   
Once the user is happy with the script, the system provides visual framing to explore the characters' visual design.
It generates a character board that contains visual details of the costumes, props, and backgrounds to distinguish the characters using a text-to-image model, DALL-E 2~\cite{dalle2}. 
Based on the character board, it generates a storyboard that shows suggested emotions, actions, and dialogs for each character. 
Users can then make reels based on these materials. 

This paper makes the following contributions:
\begin{itemize}\itemsep0em
\item Narrative framing as a structure in human-AI co-creation workflow to reframe content from one medium to another, with foundational details for executing the framing.
\item An analysis of news reel scripts that lead to three narrative framings to balance information and entertainment: expository dialog (more informative), reenactment (mixed) and comedic analogy (more entertaining).
\item A set of expert guidelines and prompting techniques for generating news reel premises, scripts, character boards, and storyboards using multimodal generative AI.
\item ReelFramer, an interactive AI workflow for translating news articles into reels by helping users produce scripts and storyboards that capture the appropriate information and tone.
\end{itemize}

We conclude with a discussion of what role humans and AI take in the co-creative process.
We also reflect on the benefits of using narrative framing and foundational details in content retargeting.
\section{Related Work}

\subsection{News Reels and Narrative Framing}
News reels are an emerging trend in journalism to help news outlets reach younger audiences who get their news almost exclusively from social media and, increasingly, from short-form videos on TikTok and Instagram~\cite{newman_news_tt_reuters_oxford}.
A 2022 report from the Reuters Institute found that 49\% of top news publishers worldwide are regularly publishing content on TikTok, most of whom started in 2021 or later~\cite{newman_digital_news_2022}. 
Additionally, they found 15\% of 18-24-year-olds already use TikTok for news~\cite{newman_digital_news_2022}, demonstrating there is an appetite for news content in this new format.

In interviews conducted by the Reuters Institute in 2022 with news reel creators~\cite{newman_news_tt_reuters_oxford}, many different approaches to creating reels have been found, but a key challenge is to balance between information and entertainment~\cite{savolainen2022infotainment}. 
If a serious topic is portrayed too light-heartedly, it may diminish the perceived seriousness of the issue~\cite{davis2022infotainment}.
However, if a video does not grab an audience's attention with something that is interesting and entertaining, they may scroll past it.
Some news organizations take a conservative approach and stay closer to the TV journalism format, but others lean into the tropes of social media videos to blend news information into an engaging social media style.
The challenge is to help journalists translate their articles into this style.

Narrative framing guides the overall presentation of a story and shapes the comprehension of the represented events~\cite{schmid2021narrative}.
In literature, writers utilize narrative framing to structure the context, perspective, and overall understanding of their story~\cite{richardson1990writing}.
In journalism, narrative framing involves the strategic use of language and the organization of information to create a particular interpretation or reaction among the news audience ~\cite{hicks2016writing, wasike2013framing}.
For example, to achieve a blunt and serious tone that emphasizes facts, journalists often use an inverted pyramid narrative frame~\cite{pyramid} to structure their story, presenting the most crucial information---such as who, what, when, and where---first, followed by details arranged by decreasing importance~\cite{pyramid}. 
However, to convey a different tone, a different narrative framing would be more appropriate.

To construct journalistic narratives, the concepts of news value, news angle and news frame offer distinct considerations. 
News value helps journalists prioritize which events to report~\cite{harcup2017news}.
For example, a timely event with conflict would typically have high news value and high priority to report on. 
News angle concerns the selection of the specific aspect of the event being reported~\cite{motta2020analysis, anglekindling}. 
For example, a press release about a wind farm can highlight either economic development or environmental consequences as its angle.
News frame is a concept in the field of communication that reflects journalists' perceptions of their role in society and, therefore, the intended outcome of their articles~\cite{valkenburg1999effects, valenzuela2017behavioral}. 
For example, the goal of holding the powerful to account may result in framing a story as a conflict between a powerful person and other interests.
Print articles have already examined the news value of an event and deemed them to be worthy of publication. 
In the process of writing the article, the journalists have embedded the news angle that is appropriate for the event and publication.
News frame is not necessarily consciously chosen by or even evident to journalists, but it is also determined during the writing of the news story.
Thus, when making a reel from a print article, the news value has already been established, the news angle and news frame have been embedded---reels should preserve these choices rather than change them. 

Adapting print articles to the reel format requires a creative reimagining of their presentation to meet the demand for engaging storytelling on social media.
Narrative reframing can help achieve this goal.
In contrast to traditional narrative framings used in journalism like the inverted pyramid, news reels often aim to achieve a personal, engaging, and lighthearted tone consistent with social media in a short length. 
However, facts and information still need to be conveyed in a credible and respectful way.
Thus, we explore new narrative framings in ReelFramer for journalists to translate print articles into social media reels for new audiences.

\subsection{Automated Narrative Generation}
Automated storytelling is an AI task that aims to create coherent and imaginative narratives. 
Many early explorations revolved around AI planning~\cite{weld1999recent}. 
For example, a partial order causal link (POCL) planner was developed to generate plot progression by creating character goal hierarchies and intentions~\cite{riedl2010narrative}.
Similarly, a refinement of the POCL planner was used for the adaptation of game plotlines~\cite{li2010offline}.
While existing work has mainly focused on fiction writing, script-based methods~\cite{schank1977scripts} have been used to automate limited types of journalistic stories, especially those driven by data, such as financial earnings reports and sports articles~\cite{ap_automated_insights, caswell2018automated}. 

Neural network-based language models like Seq2Seq~\cite{sutskever2014sequence} can generate stories without explicit planning.
However, these models often generate incoherent stories with unclear direction because they only consider preceding tokens without a strategic plan for subsequent content~\cite{daniluk2017frustratingly}. 
To address this, several solutions have been proposed. 
One method involves outlining key story points as guides to fill in the narrative~\cite{WANG2023126792}. 
For instance, the causal and commonsense plot ordering framework~\cite{ammanabrolu2021automated} produces narratives through plot infilling with soft causal relations derived from a knowledge graph~\cite{comet}. 
Another strategy conditions story generation on predefined plot points~\cite{WANG2023126792}. 
Using a plan-and-write approach, explicit storyline planning results in stories that are more diverse, coherent, and on-topic~\cite{yao2019plan}.
However, producing a sequence of coherent and logically consistent sentences remains an open challenge~\cite{yao2019plan}.

Recently, LLMs have demonstrated unprecedented abilities in story generation~\cite{xie2023next}.  
LLMs can generate stories from a single prompt, but the outputs are far from perfect.
LLM-generated stories currently pass 3-10 times fewer tests of creative writing than those written by professionals, across the dimensions of fluency, flexibility, originality, and elaboration~\cite{chakrabarty2023art}.
It is thus important to incorporate humans in the process to guide and control the generation. 

\subsection{Computational Storyboarding}
Storyboarding is an essential part of developing films and other visual story media. 
Storyboards allow creators to visualize the main events through a series of images and identify holes or inconsistencies in the narrative before filming begins. 
They are also essential for planning shots and determining the actors, props, and settings that will be needed. 
They are challenging to create computationally because they require consistent image generation across multiple scenes. 
While technologies like DALL-E~\cite{dalle2}, Midjourney~\cite{midjourney}, and Stable Diffusion~\cite{stable_diffusion} are impressive, they are difficult to control, although progress has been made towards more consistent images ~\cite{neural_storyboards, maharana2022storydall} and producing changing facial expressions on otherwise static characters~\cite{xiaojuan_expressions}. 
While there have been works in supporting stereographic storyboarding~\cite{Storeoboard} and creating storyboards based in code~\cite{CodeToon}, this is an area that has many future directions to explore.

\subsection{Human-AI Co-creativity}
Generative AI has already been hugely successful in supporting creative tasks, including music composition~\cite{cococo}, visual design~\cite{opal, popblends}, and writing~\cite{sparks,talebrush,SAGA}. 
LLMs, in particular, are transforming various creative endeavors.
Trained on billions of documents, LLMs contain a vast amount of general world knowledge and can perform numerous NLP tasks without pre-training~\cite{gpt,language_models}.
LLMs have been used as open-ended collaborative writing tools to explore different styles~\cite{wordcraft}, ideate engaging tweets~\cite{tweetorial_hook}, suggest reader perspectives~\cite{sparks}, and find angles for news stories~\cite{anglekindling}. 

However, generative AI still comes with challenges. 
It is known to hallucinate and provide false information~\cite{gpt}. 
Many tasks are too complex to accomplish in one prompt and have to be broken into smaller problems, then chained together ~\cite{cai_ai_chains,promptchainer,promptmaker}. 
Additionally, generative AI is difficult to control. 
Prompting with text is the main interface for text, image, and code generation, but this makes it hard to refine or iterate on outputs.
Human involvement is essential for guiding, correcting, and performing tasks that generative AI cannot complete independently.

Many human-AI co-creation systems leverage structured prompting to allow users to guide and control the generation.
Structured prompting involves providing LLMs with a templated input to guide them toward producing a high-quality output. 
For example, BotDesigner~\cite{zamfirescu2023johnny} offers a prompt template that includes preambles, first turns, and reminders to aid non-experts in creating instructional chatbots for tasks like following a recipe.
Calypso~\cite{zhu2023calypso} is a chatbot to help D\&D Dungeon Masters craft narratives during game play. 
It structures its prompting by first generating a summary of encounter statistics before creating a chatbot for users to engage with.
Similarly, Dramatron~\cite{mirowski2022cowriting} uses structured prompting to assist screenwriters in expanding plot summaries (known as log lines) into complete long-form scripts. 
Its structured generation approach first creates descriptions of characters and settings, then develops the plot into scenes following popular story structures, such as the hero's journey. 
This works well for expanding a plot summary into well-established story structures, however, reframing news comes with different challenges. 
To retarget news into reels, we have to engage users in a structured exploration of the news to find a narrative framing and foundational details that fit the story (rather than starting with one already established). 
Therefore, we provide narrative framing as a structure in ReelFramer to facilitate such exploration.
Additionally, since new reels must maintain key information, ReelFramer also assists users in determining which information to convey as part of the narrative details.
This is a key constraint in accurately conveying facts. 
\section{Co-Design with Journalists}
To build a tool that would be useful to journalists, we engaged in a six-month co-design process with two professional journalists: one specializing in television and the other with expertise in media innovation. 
Both have extensive experience in producing journalism and educating young journalists at the college level.
They actively participated throughout the design process, from understanding the problem and identifying needs to brainstorming, prototyping, and testing. 
They also experimented with the prototypes, providing feedback that helped iterate the design to ensure it aligned with journalistic requirements.
We met with the journalists for an hour each week to analyze existing news reels and form guidelines on what they both deemed to have journalistic value. 

\subsection{Types of News Reels to Support}
Existing news reels take on many forms---some are purely informational, some are purely for entertainment (like reaction videos), and others are in between with elements of social media reels interwoven with news content. 
While these can be fun and sometimes appropriate, they are not the focus of this work.
Although our co-designers did not always share the same taste, they both liked the compromise between information and entertainment in the roleplay-style news reels, where an actor portrayed two or more characters from the news article and acted out a conversation between them to convey the news, and some humorous quips, costumes, props, and backgrounds. 
They particularly liked the style of \textit{The Washington Post}, which balanced between having lighthearted introductions (such as treating Hurricane Ian like a tourist traveling to Florida) and conveying serious information (such as describing the extensive damage hurricanes had recently inflicted). 
These videos get hundreds of thousands and sometimes millions of views on TikTok, which speaks to their effectiveness in reaching audiences.  
Additionally, roleplay-style reels are naturally suited to TikTok. 
Compared to reels that resemble TV news segments~\cite{newman_news_tt_reuters_oxford}, our co-designers viewed the roleplay format as a better fit for the platform.
Thus, we decided that \textbf{creating roleplay-style news reels would be a valuable goal to support}.

\subsection{Target User Group}
Since news reels are an emerging trend, there is not a set journalistic role for creating them. 
Currently, the people in news organizations making news reels are often recent graduates who are steeped in social media and are given the freedom to create in their own style
~\cite{Pompeo_2023}. 
Thus, \textbf{the target audience of our tool is recent journalism graduates}---this group has a journalism background and understands the values and responsibilities of journalists, but also appreciates the immense role social media plays in the lives of young people. 
These are the people most likely to step into the role of creating news reels. 
Because this job does not formally exist in many organizations, there is no explicit training for it.
Thus, our co-designers thought it would be immensely valuable to have tools that support young journalists in creating their first news reels.
They even considered using such tools in class as a teaching tool for scaffolding the process of creating news reels.

\subsection{Creation Process to Support}
The co-designers agreed that roleplay-style news reels are best when acted out by a real person.
Compared with animation-style reels, reels with a real person show more authenticity and engage the audience on a more personal level~\cite{newman_news_tt_reuters_oxford}.
Thus, the system should not produce a full video directly, instead, \textbf{the system should produce a script and storyboard that journalists could act out and film}.
It is also expected that there will be many rounds of ideation and iterations before a final script or storyboard is produced. 
Thus, an interactive system, like a workflow, would be ideal.
Currently, almost all news reels are based on an existing news article that has already been researched, written, and published in the traditional print media style. 
Thus, the challenge is to transform a traditional news article into the news reel format.

\subsection{Developing Narrative Framings} 
When analyzing reel examples to develop narrative framings, our co-designers stressed that in journalism, there are many ways to tell a story.
Even within the space of roleplay reels, there are types of roleplay that focus more on information or more on entertainment. 
To understand the variety within roleplay-style reels, our co-design team analyzed 50 such reels from \textit{The Washington Post}, \textit{LA Times}, and \textit{The Wall Street Journal}. 
For each reel, we examined its narrative structure and linked it back to the original article. 
Our co-designers then assessed whether the reel effectively communicated the news and adhered to journalistic norms. 
Together, we categorized the reels based on their varying degrees of informational and entertainment value. 
We decided to support three basic narrative framings that are representative of the analyzed reels and span the spectrum of infotainment: \textit{Expository dialog}, \textit{Reenactment} and \textit{Comedic analogy}. 
These framings diverge from traditional narrative framings by prioritizing a mix of information delivery and entertainment and facilitating the adaptation of news articles for social media audiences.
We provide framing examples for an article about a global shortage of chips that causes delays in credit card issuance\footnote{\url{https://www.washingtonpost.com/business/2023/02/14/credit-card-delays/}\label{credit-card-article}}.

\FloatBarrier
\begin{table*}[t]
\begin{center}
\renewcommand{\arraystretch}{1.1}
\begin{tabular}{|p{0.2cm}|p{2.4cm}|p{6.4cm}|p{6.2cm}|}
\hline
\multicolumn{2}{|c|}{\textbf{Narrative Framing}} & \multicolumn{1}{c|}{\textbf{Narrative Details (Premise)}} & \multicolumn{1}{c|}{\textbf{Script}}  \\ \hline
\multirow{11}{*}{\rotatebox[origin=c]{90}{\textbf{Expository Dialog}}} & \textbf{Characters} \newline \textit{(An expert and a newcomer)} & Supply chain experts (like professor Patrick Penfield, expert) and Consumers (like Ed Delaney, naive newcomer) & {\multirow{10}{20em}{\textbf{Patrick Penfield}: Ed, you ever heard about the global chip shortage? \newline
\textbf{Ed Delaney}: Ah, those lil' shiny squares in credit cards, yeah? \newline
\textbf{Patrick Penfield}: (Smiles) Yep, that's them. Demand's skyrocketed during the pandemic, causing major delays in delivering credit and debit cards. \newline
\textbf{Ed Delaney}: Can't we just make more? I mean, we're in a chip factory! \newline 
\textbf{Patrick Penfield}: Wish it was that simple, Ed. These little fellas take 20 to 25 weeks---it used to be just 10 to 14 weeks. The pandemic has led to a surge in demand across all sectors---automotive, consumer electronics, you name it...}}   \\
\cline{2-3} 
& \textbf{Plot} \newline \textit{(Expert explaining info to newcomer)} & Professor Patrick Penfied explaining the information to Ed Delaney & \\
\cline{2-3} 
& \textbf{Setting} \newline \textit{(News-related)} & A chip manufacturer's factory &  \\ 
\cline{2-3} 
& \textbf{Key information} \newline \textit{(Three main information points)} & - The global chip shortage is causing major delays in delivering credit and debit cards. \newline
- Semiconductor demand has spiked during the pandemic, leading to long lead times for chip production, averaging 20-25 weeks. \newline
- Competition from other industries has kicked the credit card sector to the bottom of the priority pile.
&   \\ 
\hline
\hline

\multirow{11}{*}{\rotatebox[origin=c]{90}{\textbf{Reenactment} }} & \textbf{Characters} \newline \textit{(Key stakeholders)}  & Credit and debit card issuers (particularly credit unions) and Consumers (like Ed Delaney) & {\multirow{10}{20em}{\textbf{Ed Delaney:}
Excuse me, where's my new card? It's been 6 weeks! \newline
\textbf{Credit Union:}
(scratches head) Well, Mr. Delaney, we're experiencing a global chip shortage. \newline
\textbf{Ed Delaney:}
(incredulous) A global shortage of potato chips? What's that got to do with my card? \newline
\textbf{Credit Union:}
(laughs) No, not potato chips! Semiconductor chips for the cards. It's out of our control... The demand for these chips is too high. \newline
\textbf{Ed Delaney:}
(sarcastic) Oh, sure, blame it on the high demand for these invisible chips. \newline
\textbf{Credit Union:}
(smiling) Well, it's not just us. All industries are struggling with limited factory production...}}   \\
\cline{2-3} 
& \textbf{Plot} \newline \textit{(Characters acting out what happens in real time)} & 
The global chip shortage is affecting credit and debit card issuers, particularly credit unions, causing delays of weeks or even months in consumer card delivery times & \\
\cline{2-3} 
& \textbf{Setting}  \newline \textit{(News-related)} & A credit union office  &  \\ 
\cline{2-3} 
& \textbf{Key information} \newline \textit{(Three main information points)} & - The global chip shortage is causing major delays in delivering credit and debit cards. \newline
- Semiconductor demand has spiked during the pandemic, leading to long lead times for chip production, averaging 20-25 weeks. \newline
- Competition from other industries has kicked the credit card sector to the bottom of the priority pile.
&   \\ 

\hline
\hline

\multirow{15}{*}{\rotatebox[origin=c]{90}{\textbf{Comedic Analogy }}} & \textbf{Characters} \newline \textit{(Key stakeholders)}  & Credit and debit card issuers (particularly credit unions) and Consumers (like Ed Delaney) & {\multirow{9}{20em}{\textbf{Ed Delaney:}
Excuse me, when can I expect my cookies? \newline
\textbf{Credit Union:}
Oh, we're working on it, but we're short on chips! You won't get your card---I mean cookies---for at least six weeks! \newline
\textbf{Ed Delaney:}
(confused) Six weeks? But I'm hungry now!  It used to be just five to ten days! \newline
\textbf{Credit Union:}
We hear you, but we're swamped with orders! \newline
\textbf{Ed Delaney:}
This is ridiculous! I'm starving! \newline
\textbf{Credit Union:}
(nervously) We know, but we got caught up in a perfect storm. High demand, limited production facilities, and industry competition. \newline
\textbf{Ed Delaney:}
Just our luck. Stuck behind a bunch of other industries, waiting for our ``chips.''  \newline
\textbf{Credit Union:}
(apologetically) We're sorry, but you'll have to wait like everyone else...}} \\

\cline{2-3} 
& \textbf{Plot}  \newline \textit{(Comedic analogy)} & The credit union is like the pastry chef, and consumers are hungry customers waiting in line for chip-enabled cookies & \\
\cline{2-3} 
& \textbf{Setting} \newline \textit{(Analogy-related)} & A busy bakery &  \\ 
\cline{2-3} 
& \textbf{Key information} \newline \textit{(Four major plot points)} & - The global chip shortage is causing delays in issuing debit and credit cards, with some consumers experiencing wait times of six weeks or more. \newline
- Credit union members seem to be particularly affected, with typical time for card issuance stretching from five to 10 days to weeks or even months. \newline
- Experts predicted that the delays in card deliveries will continue throughout 2023 despite projections of 3 billion cards being manufactured this year. \newline
- The chip shortage results from high demand, limited production facilities, and competition... &   \\ 
\hline
\end{tabular}
\end{center}
\caption{Narrative framings, narrative details (or premise) and script examples for a news article about the global shortage of chips that causes delays in credit card issuance\footref{credit-card-article}.}
\label{table:framings}
\end{table*}

\textbf{1. Expository dialog}. 
This framing focuses more on information than entertainment. The plot is that an expert on the topic explains the situation to a naive newcomer. 
This newcomer asks basic questions to understand the topic. 
The expert/newcomer trope is widely used in narratives of all forms. 
For news reels, the characters are taken from the story itself. 
For example, the expert could be a professor of supply chain and the naive newcomer could be a customer who is trying to get a credit card issued. 
The professor introduces the global chip shortage, and the newcomer asks why, allowing the professor to explain the reasons and consequences of this phenomenon. 
This writing technique is often called an ``infodump'' because it is an easy and efficient way to convey information.

\textbf{2. Reenactment}. 
This framing has aspects of expository dialog but has more entertainment value. 
Instead of explaining past events, the plot involves the characters acting out the event in real time and discovering the consequences. 
For example, one character is the customer who is trying to get a credit card issued.
The other character is the credit union, explaining that card issuance will be delayed because there is a global shortage of chips, and the credit card sector is at the bottom of the priority pile. 
Reenactments are entertaining because ``acting it out'' is a way to heighten the emotions, stakes, and humor in a story~\cite{humor_Holloway2010,humor_Carter2001}.

\textbf{3. Comedic analogy}. This framing primarily focuses on comedy, and the final product strays further from the original story. 
The plot is an activity that is analogous to the real event. 
For example, the plot can be the credit union serving as the pastry chef and its consumers as the hungry customers waiting in line for chip-enabled cookies. 
The customers complain about why it takes so long to get the cookies like they have to wait for the chip-enabled credit cards.
Analogy is also a classic technique in comedy and drama, but it is challenging because it requires finding a vehicle that is appropriate to the tone and information of the story~\cite{humor_Raskin2009,humor_Dean2000}. 

There are many other possible framings, and there are even ways to blend or combine framings into new framings. 
But these three framings encompass major trends in news reels we analyzed and have different trade-offs in the infotainment space. 
In the Limitations section, we discuss additional framings to explore, including ones that use trending songs and visual gags.

\subsection{Developing Scripts} 
\label{section:premise_parameters}
Generative AI opens up the possibility of generating novel text based on simple instructions. 
There is obvious potential to generate scripts and screenplays, which has been explored before~\cite{mirowski2022cowriting}. 
However, the technology is far from achieving perfect outputs. 
In early prototyping sessions, our co-design team encountered many challenges in generating reel-style scripts. 
When GPT-4 is asked to produce a script, it seems to create a Hollywood-style script that does not fit the reel style. 
It often has a title, multiple characters, complex stage directions, and most problematically, a happy ending. 
Very few news stories have a true ``ending'' for their characters, and seldom is it a happy Hollywood-style one. 
To overcome this, we have to write specific directions for LLMs to adapt.

\textbf{Script style parameters} are parameters that are typical of reel-style scripts, differing from Hollywood-style scripts:
\begin{itemize}\itemsep0em
    \item \textbf{the overall length} of the dialog is typically 10-12 lines long;
    \item \textbf{each line of dialog} is short---less than 20 words. The characters take turns to speak in short clips. There are no long monologues, which is more typical of info-dumps in long-form videos such as film;
    \item \textbf{the tone of the dialog} is colloquial and engaging;
    \item \textbf{the ending} should contain a punchline: a joke, concluding thought, or takeaway rather than a happy ending.
\end{itemize}

\textbf{Script content parameters} are parameters specific to the news article that need to be included in the script:
\begin{itemize}\itemsep0em
    \item \textbf{characters} should be taken from or derived from the article;
    \item \textbf{the plot} must conform to the selected narrative framing;
    \item \textbf{the setting} must be relevant to the article;
    \item \textbf{the information} must incorporate key facts or takeaways from the news article.
    Since reels are so short, they do not contain all the facts; they focus on conveying four or fewer salient pieces of information.
\end{itemize}

However, script generations were still somewhat wild and often incoherent. 
The biggest problem for the co-designers was that the original news content was often ignored or incorrect.

To help ensure the script contained correct and sufficient coverage of news, we introduced an intermediate step to the co-design process. 
In this step, the LLM would suggest the three most important news points, and the journalists could review, regenerate, or rewrite them. 
Once the journalists were happy with the news points, they were fed as an additional requirement to the LLM prompt for script generations. 
The journalists agreed that this tended to improve the quality of the script news content. 
They also requested the ability to suggest and finalize the characters, plot, and setting to feed into the script prompt. 
We realized that by establishing foundational details of the narrative framing (see Table \ref{table:framings}), the authors felt the scripts were more coherent and informational.
They also enjoyed having more control over the process. 
Overall, having narrative framings and their foundational details enabled a better co-creative process between human and AI.

\subsection{Developing Storyboards} 
Scripts are an essential starting point of a video, but storyboards are also necessary to help think through the visual aspects of the story.
For news reels, there are several challenges:
If one actor plays two characters, how will the viewers tell them apart?
What clothing, props, stance, background, and other visual cues can help distinguish the characters?
Do the visual aspects of the story hit the right beats?  
Since each line of dialog is a cut, are the cuts (and lines of dialog) of appropriate length?
These are difficult to assess in the script alone but can be easily evaluated in a storyboard.

Text-to-image generative models are powerful tools for creating novel imagery, but we faced challenges directly applying them to storyboards. 
Visual consistency is a major challenge. It is nearly impossible to generate the same character in different poses across different frames through prompting alone.
Being very specific about their clothing, hair, and appearance helps.  

The character board provides a foundation for the visuals in the storyboard: the characters, clothing, and props.
By establishing the detailed descriptions of characters, it allows the storyboard to generate more visually consistent depictions of the same character in each frame. 
Moreover, it allows the user to explore different ``casting decisions''---What should the characters wear? What props should the characters use? 
Since it is likely the same person plays all characters, it is important that the costumes and props help visually distinguish the characters. 
It is also important that the costumes and props are achievable by the actor.

Due to the limitations of the current text-to-image models such as DALL-E 2~\cite{dalle2}, there are still issues around the storyboard visuals.
The model struggles to portray multiple specific characters on one screen. 
It also seems biased towards complex scenes, which are inappropriate for storyboarding.
However, the model is decent at depicting poses, hand gestures, and facial expressions, which could serve as the performance guidelines for roleplay-style reels.
Luckily, TikTok storyboard tiles are fairly simple---they rarely require multiple people in one shot and do not need to be perfectly consistent to be useful for the final video recording.
It took many rounds of prompt engineering to come to a good compromise.
\section{Expert Evaluation of LLM Capabilities}
Large language models (LLMs) are excellent at producing fluent and topical text based on instructions with little or no extra training.
However, a known weakness is their ability to produce accurate content, and they often struggle with coherency in longer-form pieces. 
When prompting LLMs to write scripts based on a given narrative framing, we needed to decide how much human direction is needed.
To evaluate this, we ran a controlled study where journalists evaluated scripts generated by LLMs in both conditions.

\begin{itemize}
\item Without premise: General LLM instructions. 
\item With premise: General LLM instructions and foundational narrative details (also known as premise).
\end{itemize}

We hypothesize that scripts generated with a premise will exceed in six dimensions: 1) better conform to the narrative framing, 2) have higher information accuracy, 3) cover more important information, 4) be more coherent, 5) be more fun/entertaining, 6) better fit the TikTok style. 

\subsection{Data Preparation}
Overall, we generated 24 scripts, 12 with the premise and 12 without the premise. 
These scripts spanned six news articles and three narrative framings. 
The articles were selected from the category pages of the Associated Press website\footnote{\url{https://apnews.com/}} (U.S., World, Politics, Business, Science, Climate) in September 2023. 
Each article had four generations: two with the premise and two without, with the same narrative framing. 
Table \ref{table:generations} shows the articles, narrative framings, and experimental conditions for the 24 scripts. 

\begin{table}[t]
\begin{tabular}{| c | c | c |}
\hline
 \textbf{} & \textbf{Without premise} & \textbf{With premise}  \\
\hline
\href{https://apnews.com/article/labor-travel-weekend-travel-forecast-afb55b5995f4486cd3982aadff9a9992}{\textbf{Article 1}} & 2 Expository dialogs  & 2 Expository dialogs \\ \hline
\href{https://apnews.com/article/india-spacecraft-launch-sun-mission-aditva-isro-19b9c94c42d0e4f0cd515a8fbe459cd9}{\textbf{Article 2}}  & 2 Expository dialogs  & 2 Expository dialogs \\ \hline
\href{https://apnews.com/article/greece-archaeology-temple-ancient-offerings-kythnos-island-238f05ae4944f5ed2f5356386f076136}{\textbf{Article 3}} & 2 Reenactments  & 2 Reenactments \\ \hline
\href{https://apnews.com/article/pakistan-strike-inflation-energy-354ba19e95447d1aa7f938f0d6042c5c}{\textbf{Article 4}} & 2 Reenactments  & 2 Reenactments \\ \hline
\href{https://apnews.com/article/biden-climate-change-arizona-new-mexico-utah-94f16d6c6204377e80be4cace48bd379}{\textbf{Article 5}} & 2 Comedic analogies  & 2 Comedic analogies  \\ \hline
\href{https://apnews.com/article/new-york-city-united-states-insects-48a4bbb073aacc95e212f82a7b110d7b}{\textbf{Article 6}} & 2 Comedic analogies  & 2 Comedic analogies  \\ \hline
\end{tabular}
\caption{Articles, narrative framings and experimental conditions for the 24 script generations. }
\label{table:generations}
\end{table}

\begin{table*}[t]
\renewcommand{\arraystretch}{1.05}
\begin{tabular}{| l | c | c | c |}
\hline
 \textbf{} & Without premise & With premise & p-value \\
\hline
Q1: How well does the script 
conform to the narrative framing?
 & 5.58 (1.89) & 6.95 (0.14)  & \textbf{0.028}  \\ \hline
Q2: Is the information embedded
in the script correct?
 & 6.67 (0.55) & 6.83 (0.55)  & 0.165  \\ \hline
Q3: Does the script cover the important information in the article?
 & 4.63 (1.23)  & 5.50 (1.43)  & \textbf{0.034} \\ \hline
Q4: How coherent is the script?
Can you understand what happens? & 4.63 (1.46) & 5.92 (0.79) & \textbf{0.011}  \\ \hline
Q5: Is the script fun/entertaining? & 4.29 (1.46) & 4.67 (1.49) & 0.236 \\ \hline
Q6: Is the script in TikTok style? & 6.50 (0.65) & 6.51 (0.54) & 0.237 \\ \hline
\end{tabular}
\caption{Evaluation results comparing without-premise and with-premise conditions, where means, standard deviations (in parentheses), and p-values for the paired-sample Wilcoxon tests are reported. Bolded p-values are statistically significant. }
\label{table:annotation}
\end{table*}

To generate scripts in each narrative framing, we used GPT-4 prompted with requirements derived from script style and content parameters in Section \ref{section:premise_parameters}. 
Both conditions had the same prompt for style parameters: ``It should be entertaining. The dialogue should be colloquial and engaging. The dialogue should be 10 to 12 lines long. Each line of dialogue should be short---less than 20 words. End it with a punchline.'' 
Both conditions start with the basic directive: ``Write a script for a comedy skit…''
In the without-premise condition, instructions are generic, like ``The characters should be taken from the article or derived from it.'' 
In the with-premise condition, details are given like ``The characters should be exactly the customer and the Credit Union.''
The prompts are similar in all possible ways, other than the with-premise condition has article-based premise details and the without-premise condition has a generic version of the content requirement.

We need to establish a coherent premise before generating scripts in the with-premise condition. 
The paper's second author finalized premises based on the LLM suggestions. 
The goal was to verify that the characters were relevant to the roles in the framing, that the information points were accurate and salient, and, particularly for comedic analogy, that the plot was coherent. 
9 of 12 premises were used unaltered from their generated state. 
3 of 12 premises were minimally edited. 
(Two of three edited premises involved changing a character to better fit the roles in the framing, while the final one involved removing extraneous words in the generated plot.)

\subsection{Participants and Procedure}
We invited four experts (different from our co-designers) to evaluate the scripts. 
The experts (2 female, 2 male; average age 22.0) are all from a journalism background (either having studied journalism, worked at a newspaper, or earned journalism awards) and actively engaged with TikTok for at least three hours daily. 
Before the evaluation, we presented experts with examples of the three narrative framings. 
We provided an evaluation rubric of the six dimensions and guided them through it using practical examples. 
We requested them to read the original articles thoroughly before commencing ratings. 
For each of the 24 scripts, two experts independently evaluated each dimension on a 7-point scale with justifications.
The experts were compensated \$50/hour. 

\subsection{Evaluation Results and Findings}
The inter-rater reliability shows substantial agreement between the two experts' judgments with Cohen’s $\kappa = .71$.  
This is deemed acceptable due to its highly subjective nature. 
We average the two experts' scores for each script generation and dimension. 
We run paired-sample Wilcoxon tests to compare the scores (see Table \ref{table:annotation}).

Overall, the with-premise condition outperformed in all dimensions. 
Particularly, the with-premise condition outperformed significantly in the following three dimensions: conforming to the framing ($p$=0.028), covering important information ($p$=0.034), and coherency of the script ($p$=0.011). 
In all other dimensions, the with-premise scores were higher but not as significantly. 

\textbf{Scripts with a premise better conform to the narrative framing.}
In the with-premise condition, the LLM is explicitly given the key script elements that are ``guaranteed'' to conform to the framing.
This restricts the scripts from deviating from the core concept, particularly in comedic analogy scripts that require complex plots. 
The central analogy must be clearly defined to succeed, which is difficult to achieve without the intermediate premise step. 
For example, in a without-premise generation for \href{https://apnews.com/article/biden-climate-change-arizona-new-mexico-utah-94f16d6c6204377e80be4cace48bd379}{Article 5}, a story about Biden's climate-forwarding campaign in the Western US, the LLM fails to identify a central analogy. 
Instead, it makes many ``comedic'' comparisons that are irrelevant to the article. 
As a result, the article's central idea is not captured by an overarching metaphor. 
Rather, it is completely lost, making the script a poor example of a cohesive comedic analogy (score: 4.0/7). 
In comparison, a with-premise generation centers around a single metaphor verified in the premise: Biden's journey to the West Coast is like an old Western cowboy fighting for justice. 
Every action and dialogue in the script emphasizes this one apt analogy, effectively expressing the article's main theme through an amusing metaphor (score: 6.5/7).

\textbf{Scripts with a premise cover more important information.} 
With the most important information explicitly provided in the premise, the directions for the required content are very clear, and the LLM manages to incorporate it well. 
When this information is not provided, the directions are much more nebulous: the model must synthesize, prioritize, and integrate information from the article in one step. 
This provides far more room for interpretation of the model and can result in vastly different results. 
Also, without the explicit content parameters, the stylistic demands are the most clearly defined requirements in the prompt. 
Consequently, in the without-premise case, it often covers only one point (and often not the most important one) to focus on being funny or coherent. 
For example, \href{https://apnews.com/article/india-spacecraft-launch-sun-mission-aditva-isro-19b9c94c42d0e4f0cd515a8fbe459cd9}{Article 2} highlights India's recent successes in space missions, propelling it to become a leader in international space exploration.
However, one of its without-premise generations forgoes describing these important details, instead focusing on maximizing the comedic style of the script by including many jokes. 

\textbf{Scripts with a premise are more coherent.}
A verified premise provides coherent and relevant characters, setting, information points, and plot.
Without this planning, scripts can easily become incoherent. 
The premise works similarly to a thesis, giving the final script a scaffolding of exact points to cover, a storyline to follow, and characters to use. 
It is difficult to express a series of thoughts effectively and coherently without a premise, similar to an essay without a thesis. 
This can be seen in a without-premise generation for \href{https://apnews.com/article/india-spacecraft-launch-sun-mission-aditva-isro-19b9c94c42d0e4f0cd515a8fbe459cd9}{Article 2} (Indian space exploration), an expository dialog where one character is only named as a newcomer with no further details. 
The setting, described as a space-themed cafe, is also strange and mostly irrelevant.
Because the setting and the character have a flimsy relationship with the expert character, their interaction has little precedent. 
As a result, the jokes seem occasionally nonsensical, and because there was no key information provided, the facts are poorly explained and awkwardly integrated.

\textbf{However, scripts with and without a premise scored equally for entertainment and TikTok styles.}
Entertainment and TikTok styles are stylistic elements. 
Both with- and without-premise prompts had the same language for style content. 
With no additional direction, the with-premise condition had no advantage over the without-premise condition.
Seemingly adding premise information did not give it an edge in these stylistic qualities. 
This makes sense considering the wide range of content popular on TikTok and what the users considered ``entertaining''. 
Note that both conditions do not get high scores in terms of the scripts being fun/entertaining (without-premise: 4.29/7, with-premise: 4.67/7).
This proves that the current LLMs are not particularly good at generating humor.
Using machines to generate humor still seems to be a challenging task.
In terms of the TikTok style, multiple experts stated that they found incoherent or ``random'' content amusing and characteristic of TikTok. 
As a result, the structure provided by the with-premise condition was not necessary to outperform in these fields.

\textbf{Surprisingly, scripts with and without premises scored equally for information accuracy.} 
LLMs have proven to be very successful in synthesizing information.
While in the without-premise case, it may have synthesized too much (to the exclusion of important details), this does not correlate to significantly decreased levels of accuracy.
The premise is more of the calibration step for what information should be included. 
It allows people to verify the information points before the script generation.
However, the premise does not result in information of higher accuracy, as it uses the same information resource as in the without-premise condition: the headline and the article.
Whereas for other content, such as characters, plot, etc., that ultimately forms the creative framework of the script, a creative leap is inherently needed.

Overall, the premise is a mechanism for producing better scripts, mostly by improving the content to have better information coverage and coherency and better conform to the narrative framing.
\section{ReelFramer System}

ReelFramer is an interactive system that enables users to create news reels from an article in two stages: 1) scriptwriting and 2) storyboarding. 
The process involves a human-AI collaborative workflow where generative AI makes suggestions that users can accept, edit, or regenerate. 
In the scriptwriting stage, the system introduces three narrative framings, supporting the exploration of their foundational narrative details, or premise: characters, plot, setting, and key information.
The narrative framing and premise help generate more informative and coherent scripts. 
In the storyboarding stage, the system supports visual framing through the exploration of characters and visual details. 
The system generates a character board that details the characters' appearance, props, and background, which helps generate more visually coherent storyboards.

ReelFramer is implemented in the Flask/Python web framework. 
It is powered by an LLM (GPT-4~\cite{gpt4}) for information retrieval, text understanding, and generation, a BERT-based semantic matching model (Minilm~\cite{wang2020minilm}) to identify relevant information points for script review, and a text-to-image generation model (DALLE-2~\cite{dalle2}). 
Prompts can be found in Appendix \ref{appendix:prompts}.
 
Next, we illustrate the typical interaction with the system by providing a system walk-through. 
Here, the user wants to make a news reel for an article about the global shortage of chips that causes delays in credit card issuance\footref{credit-card-article}.

\begin{figure*}
\centering
\includegraphics[width=1\textwidth]{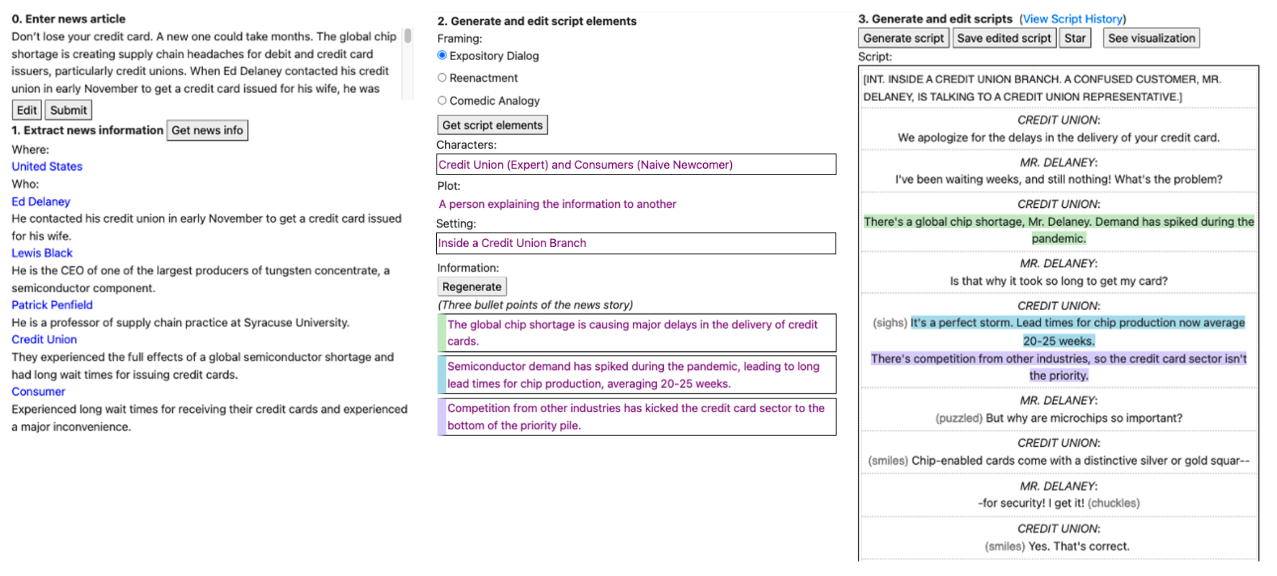}
\caption{Screenshot of ReelFramer Stage 1: Scriptwriting: 1) extract news information, 2) generate and edit script premises, and 3) generate and edit scripts. Users can regenerate suggestions in the first step, and in the subsequent two steps, they can choose to accept, regenerate, or edit the suggestions.}
\label{fig:stage1_screenshot}
\Description{A screenshot of the ReelFramer system during Stage 1: Scriptwriting. The process begins with step 0, where the user is prompted to input a news article. In step 1, the system extracts key information from the news, identifying the "where" and "who" as well as actions related to the news event. Step 2 involves generating and editing script premises, where the user selects a framing from options including expository dialogue, reenactment, or comedic analogy. Upon selection, they click "get script elements" to obtain key narrative details, including characters, premise, setting, and information points on the news story (presented in three bullet points). Moving to Step 3, users can generate scripts, save their edited versions, star favorite iterations, or visualize the script. Within the script, three lines are highlighted in different colors, each corresponding to one of the three bullet points outlined in step 2, to ensure all major news points are covered.}
\end{figure*}

\subsection{Stage 1: Scriptwriting}
Stage 1 (see Figure \ref{fig:stage1_screenshot}) aims to assist users in developing a script that is both informational and entertaining. It comprises three panels: 1) extract news information, 2) generate and edit script premises, and 3) generate and edit scripts. 

\subsubsection{Extract news information}
First, the user enters the news headline and article and submits it. 
From the news headline and text, the system uses the LLM to summarize key places, people, and actions in the story. 
It lists five important people mentioned and what they did (the ``who'' and ``what'' of the article). 
For this article, the important people include a person named Ed Delaney, who is trying to get a card issued, Credit Unions who are trying to issue new cards, experts such as a professor of supply chain management, etc. 
It also extracts the main location where the event happened. 

This panel summarizes ``who'', ``what'', and ``where'' because it is important journalistic information. 
Additionally, it aligns with the goal of developing a role-play style reel that uses people to act out events.
Users glance at it to get a sense of a person-based news summary.
They often read it more deeply when selecting the characters for the premise. 

\subsubsection{Generate and edit script premises}
Next, the user chooses one of the three narrative framings (expository dialog, reenactment, or comedic analogy) and presses the ``get script elements'' button.
Based on the framing, the system shows the user fields that need to be filled in for the premise, with suggestions automatically populated by the LLM. 
Each premise includes characters, plot, setting, and information points. 
The user can edit or regenerate LLM suggestions or write their own.

The default framing is expository dialog, often the simplest because it focuses on information rather than entertainment. 
The system encourages users to explore that first.
For this article, the characters of an expository dialog could have the credit union serve the role of an expert and the customer serve the role of a naive newcomer. 
The plot would be the expert explaining the information to the newcomer. 
The LLM also suggests a relevant setting: inside a credit union branch.
Next, the user decides what three pieces of information they want to convey. 
The LLM generally returns the correct and useful information points, but the user can regenerate points if they think it misses the main ideas or edit the points if they are inaccurate or misleading.

This panel presents the script premise that the user should agree on before going to the full content generation.
The user is encouraged to explore multiple framings, multiple premises, and multiple scripts to find one they are satisfied with. 
A good script should balance information with entertainment, convey the information and tone of the article, and the style of reels.

\subsubsection{Generate and edit scripts}
After establishing a premise, the user presses ``generate script''.
The system uses the LLM to generate the full script based on the premise, the news article, as well as the script style and content parameters (see Section \ref{section:premise_parameters}).

The script is displayed in the standard screenplay format with dialog narrow and centered and scene headings left aligned in all caps. 
The LLM does not present it this way---the system parses the LLM output to match this format for consistency and readability (see Appendix \ref{appendix:scripts} for formatting prompts).
To make it easier to evaluate the information in scripts, the system highlights each information point in the script with a corresponding color. 
When scanning the script, it is easy to see common problems. 
For example, the absence of color in the script indicates a potential omission of vital information; conversely, a long block of colored text might suggest there is too much information in one line.
This is implemented with a BERT-based similarity-matching algorithm~\cite{wang2020minilm} to find the text string most relevant to each information point that is above a minimum threshold for similarity of 0.5.
Even when using a good premise, the scripts vary in quality. 
Some are too long, some miss key information points, and some are too quirky or wild. 
Occasionally, a script is incoherent. 
It is important to be able to identify a bad script quickly and move on to a new one. 
Premises and highlighting help with this.

As always, users can regenerate, edit, or rewrite as they see fit. 
If they like the script, they can save it by clicking the ``star'' button. 
They can also edit the script and click the ``save edited script'' button to save their own version.
The user can click ``view script history'' to view all the scripts they generated, edited, or starred.

\begin{figure*}
\centering
\includegraphics[width=0.9\textwidth]{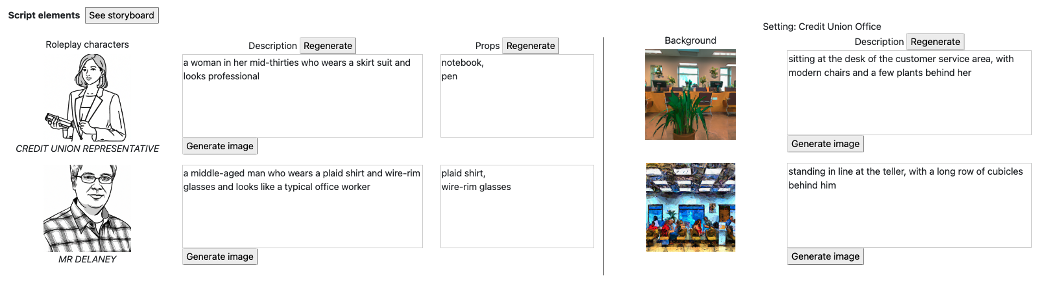}
\caption{Screenshot of ReelFramer Stage 2: Storyboarding---character board}
\label{fig:stage2_screenshot}
\Description{A screenshot of the ReelFramer system at Stage 2 Storyboard, showcasing the Character Board feature. The displayed board highlights two primary characters from a script - Mr. Delaney and a credit union representative. It features generated character descriptions, suggested props, and selectable images for each character. Additionally, the interface includes a "Generate" button that allows for the regeneration of descriptions and images. The board also proposes background descriptions alongside suggested images, assisting in the visual storyboard creation process.}
\end{figure*}

\begin{figure*}
\centering
\includegraphics[width=0.9\textwidth]{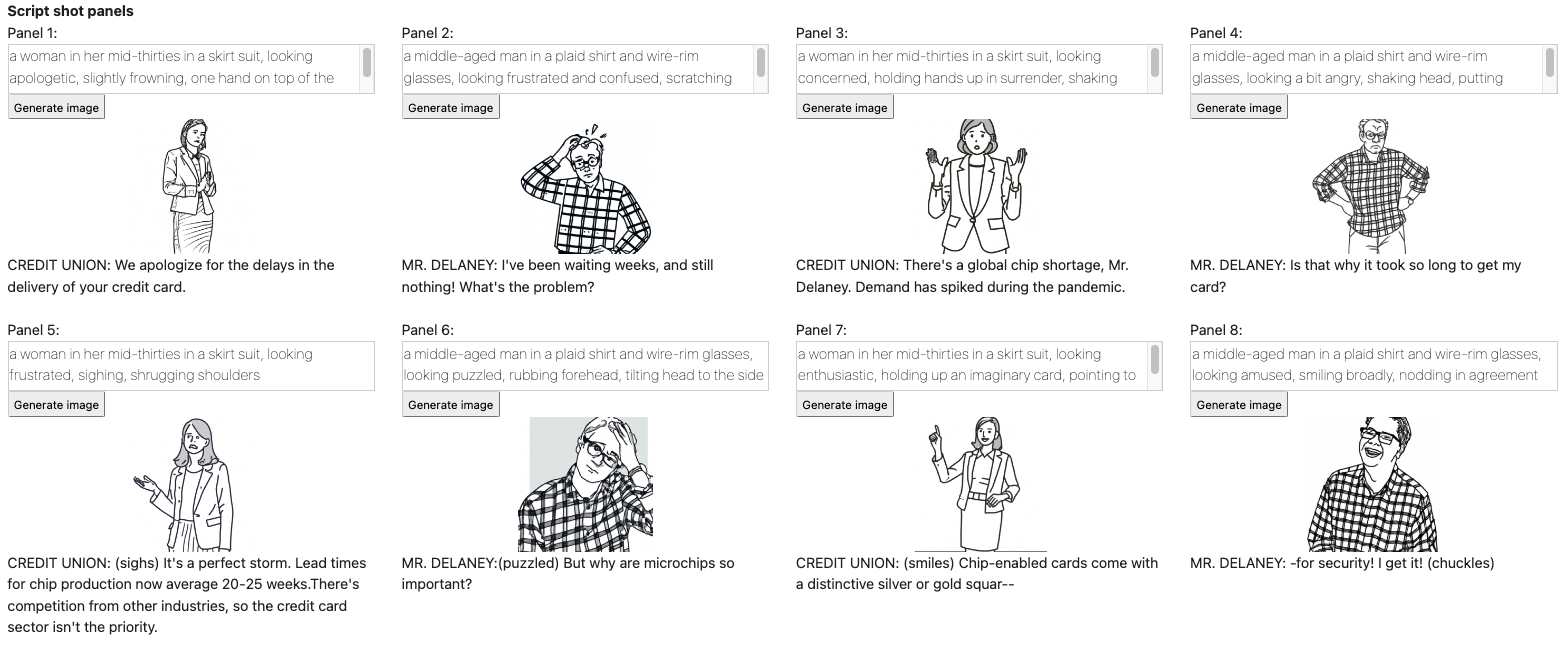}
\caption{Screenshot of ReelFramer Stage 2: Storyboarding---storyboard}
\label{fig:stage3_screenshot}
\Description{A screenshot of the ReelFramer system Stage 2: Storyboard - Storyboard. The board includes visualization suggestions for each shot, typically a dialog turn. The visualization suggestions include the character description, suggestions for expressions, gestures, and actions, as well as images. Users can edit the descriptions if they want; they can also click the "Generate" button to regenerate the images.}
\end{figure*}

\subsection{Stage 2: Storyboarding}
At the end of Stage 1, the user creates a script they are satisfied with.
The goal of Stage 2 is to help users create a storyboard to visualize the script.
It comprises two panels: 1) character board (see Figure \ref{fig:stage2_screenshot}), and 2) storyboard (see Figure \ref{fig:stage3_screenshot}).
The character board provides key visual details for generating visually coherent storyboards. 
Additionally, it allows the user to experiment with ``casting'' decisions, such as the costumes and props for the characters. 

\subsubsection{Character board} 
After finalizing the script, the user presses the ``see visualization'' button.
Based on the script and script premise, the system uses the LLM to suggest the character descriptions and a list of props to roleplay these characters.
For this article, the credit union representative can be ``a woman in her mid-thirties who wears a navy blue skirt suit and looks professional.''
To play the credit union representative, the props can be ``navy blue skirt suit, business briefcase, notebook, and pen''.
The system also uses the LLM to suggest the visual background for each character.
For this article, the background for the credit union representative can be a ``credit union office; sitting at the desk of the customer service area, with modern chairs and a few plants behind her.''
The system then uses a text-to-image model to generate the character and background images.

In this panel, the user checks the visuals of the characters and backgrounds.
If they do not like the visuals, they can regenerate the images.
They can regenerate or edit the descriptions if they do not think the content makes sense.
The user can use the list of props and character images to plan their reel recording, and they can download the background images to use directly in the reel.

\subsubsection{Storyboard} 
After establishing the character board, the user presses the ``see storyboard'' button.
The system uses the LLM to compose the shot description for each dialog turn, which contains suggested expressions, gestures, and actions.
For example, to act out the script line ``credit union: we apologize for the delays in the delivery of your credit card'', the LLM suggests ``apologetic, slightly frowning, and one hand on top of the other''.
Combining the character board and the shot description, the system uses the text-to-image model to generate the visual for each shot.

The storyboard allows users to quickly check if the script shots make sense and work together.
News reels often use simple composition, like each dialog line per cut. 
So the system uses each dialog line as the basic unit for the storyboard.
The system generates images in the simple style of ``black-and-white line art'' because it mimics the hand-drawing style that many storyboards have and makes it easier for users to reference.
If the user feels the whole story is not coming together, they can directly edit the script line or go back to iterate on the scriptwriting.
\section{User Study}
\begin{figure*}
\includegraphics[width=0.8\linewidth]{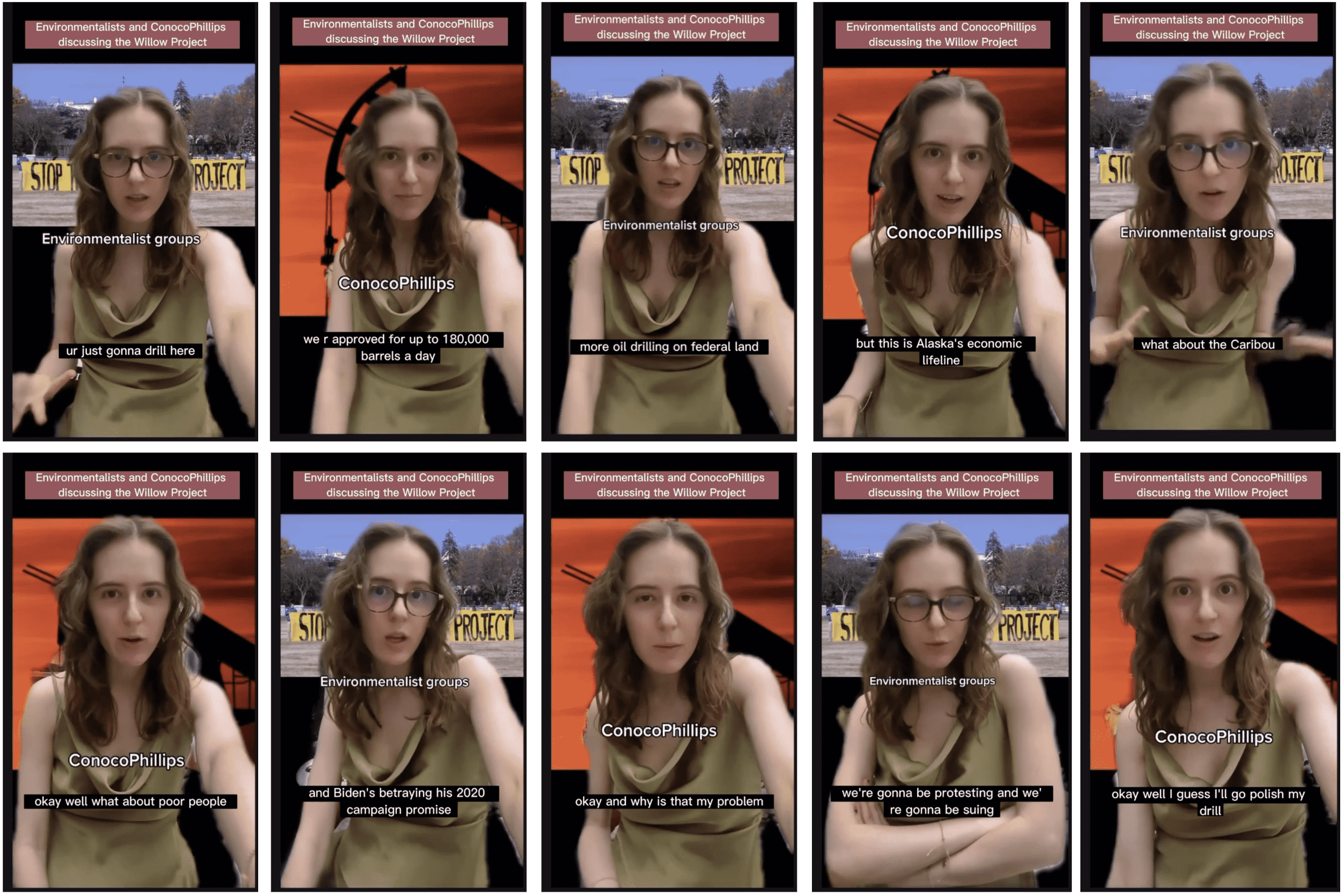}
\caption{{An expository dialog news reel created in our user study for an article about The Willow Project.}}
\label{expository}
\Description{Still screenshots of a news reel created with ReelFramer in the user study. This reel uses the expository dialog framing for a news article about ConocoPhillips drilling for oil in Alaska. The reel has a caption across each shot saying "Environmentalists and ConocoPhillips discussing the Willow project." Ten shots are captured from the reel: 1. Environmental groups: You’re just gonna drill here. 2. ConocoPhillips: We’re approved for up to 180,000 barrels a day. 3. Environmental groups: More oil drilling on federal land. 4. ConocoPhillips: But this is Alaska’s economic lifeline. 5. Environmental groups: What about caribou? 6. ConocoPhillips: Okay, well, what about poor people? 7. Environmental groups: And Biden’s betraying his 2020 campaign promise. 8. ConocoPhillips: Okay, and why is that my problem? 9. Environmental groups: We’re gonna be protesting, and we’re gonna be suing. 10. ConocoPhillips: Okay, well, I guess I’ll go polish my drill.}
\end{figure*}

\begin{figure*}
\includegraphics[width=0.8\linewidth]{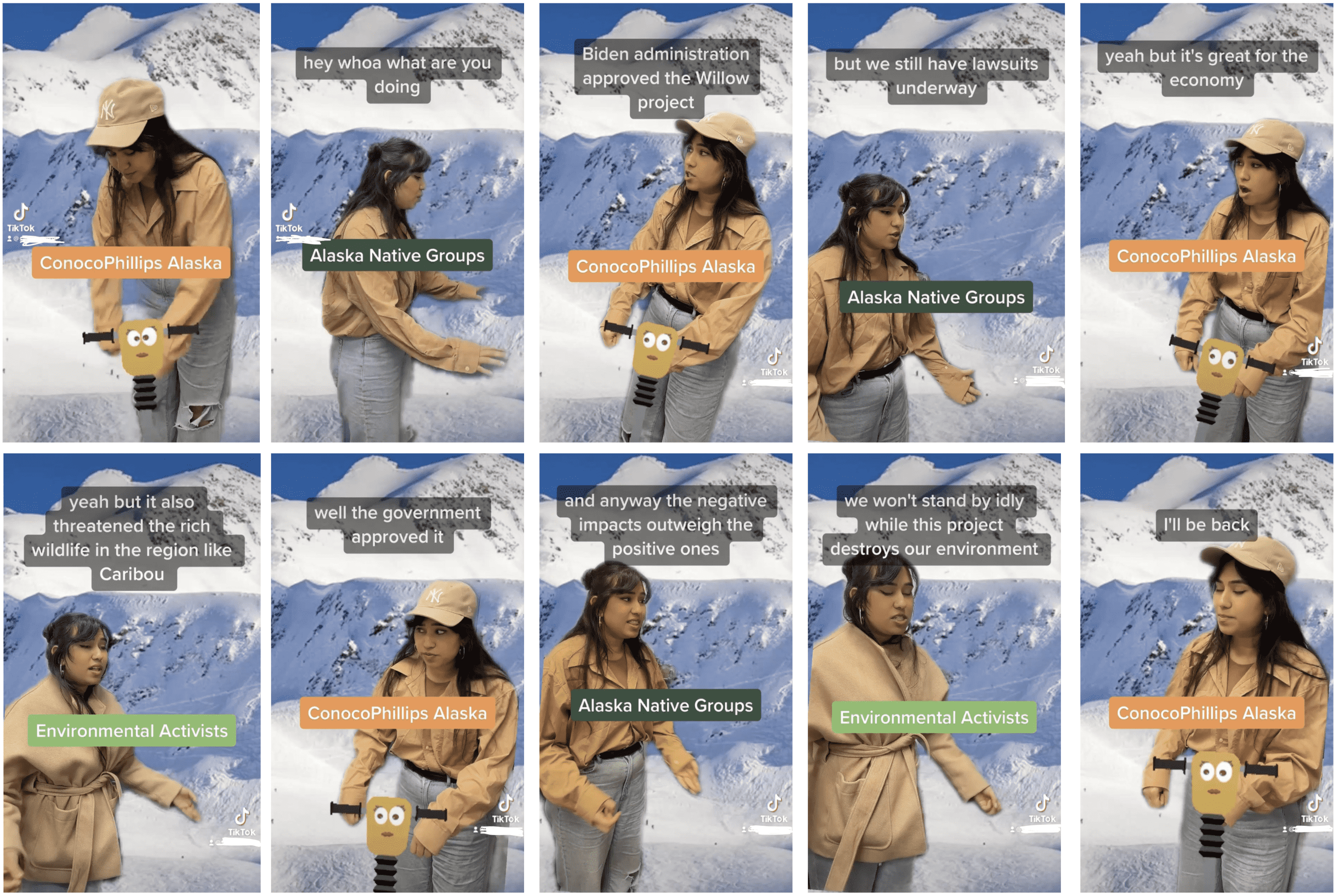}
\caption{{A reenactment news reel created in our user study for an article about The Willow Project.}}
\label{reenactment}
\Description{Still screenshots of a news reel created with ReelFramer in the user study. This reel uses the reenactment framing for a news article about ConocoPhillips drilling for oil in Alaska. Ten shots are captured from the reel: 1. [ConocoPhillips Alaska is drilling for oil.] 2. Alaska Native Group: Hey, whoa, what are you doing? 3. ConocoPhillips Alaska: Biden Administration approved the Willow Project. 4. Alaska Native Group: But we still have lawsuits underway. 5. ConocoPhillips Alaska: Yeah, but it’s great for the economy. 6. Environmental activists: Yeah, but it also threatened the rich wildlife in the region, like caribou. 7. ConocoPhillips Alaska: Well, the government approved it. 8. Alaska Native Group: And anyway, the negative impacts outweigh the positive ones. 9. Environmental activists: We won’t stand by idly while this project destroys our environment. 10. ConocoPhillips Alaska: I’ll be back!
}
\end{figure*}

\begin{figure*}
\includegraphics[width=0.8\linewidth]{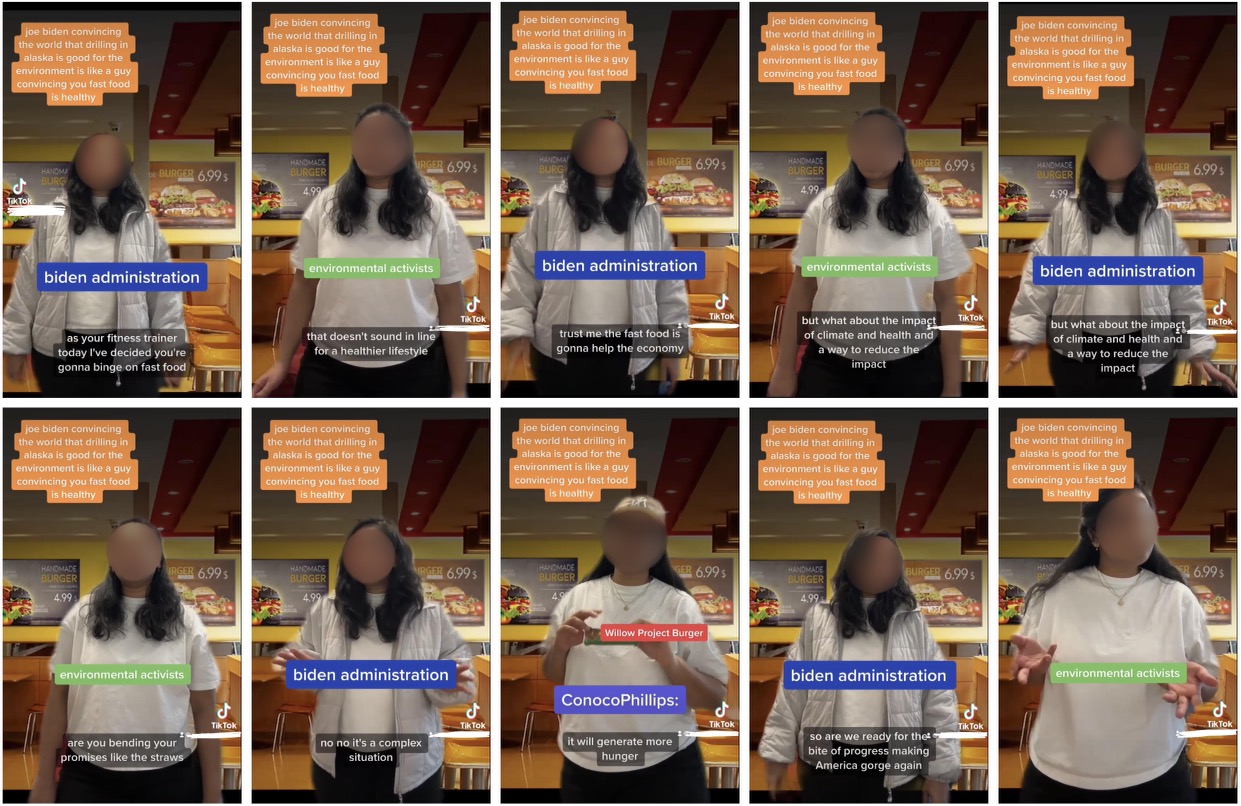}
\caption{{A comedic analogy news reel created in our user study for an article about The Willow Project. (Facial part is blurred at the request of the participant.)} }
\label{comedic}
\Description{Still screenshots of a news reel created with ReelFramer in the user study. This reel uses the comedic analogy framing for a news article about ConocoPhillips drilling for oil in Alaska. The reel has a caption across each shot saying that “Joe Biden convincing the world that drilling in Alaska is good for the environment is like a guy convincing you fast food is healthy.” Ten shots are captured from the reel: 1. Biden Administration: As your fitness trainer, today I’ve decided you’re gonna binge on fast food. 2. Environmental activists: That doesn’t sound in line for a healthier lifestyle. 3. Biden Administration: Trust me, the fast food is gonna help the economy. 4. Environmental activists: But what about the impact of climate and health and a way to reduce the impact… 5. Biden Administration: But what about the impact of climate and health and a way to reduce the impact… 6. Environmental activists: Are you bending your promises like straws? 7. Biden Administration: No, no, it’s a complex situation. 8. ConocoPhillips: Willow burger, it will generate more hunger. 9. Biden Administration: So are we ready for the bite of progress? 10. Environmental activists: Making America gorge again?}
\end{figure*}

To understand ReelFramer's effectiveness in supporting journalists in creating news reels, we conducted a user study, inviting five journalism students to each create two reels based on news articles with ReelFramer. 
Ten news reels were produced in total for further expert evaluation (see examples in Figure \ref{expository}, Figure \ref{reenactment}, and Figure \ref{comedic}).

\subsection{Participants and Procedure}
We recruited five graduate students (five female, average age 25.4, different from previous studies' participants) who were enrolled in journalism-related programs at a local university. 
All participants were active TikTok users, with four watching TikTok videos multiple times a day. 
They were also familiar with news-related TikTok videos, with all participants having seen at least two, and four having seen more than ten. 
Three participants were also avid TikTok creators: one participant had created more than ten TikToks, and two had created over 50.
However, only one participant had created a news-related TikTok before the study.
Each study session was conducted one-on-one for approximately two to three hours, and the participants were compensated \$25/hour. 

During the study, each participant's goal was to create two reels about an Associated Press (AP) article with ReelFramer. 
The participant was first briefly introduced to news reels. 
They were then provided access to ReelFramer and introduced to the system via a warm-up task---to read and create a script about \textit{Article 0}\footnote{\url{https://apnews.com/article/louisiana-wildlife-officials-illegal-pet-nutria-26dd493a2f4d7738dc59de8c4c94d850}}.
The participant was then given the option to use a pre-chosen AP article (Article 1\footnote{\url{https://apnews.com/article/banks-federal-reserve-silicon-valley-lending-fa9cea4bf6a24f881337630acf04ecfd}}, Article 2\footnote{\url{https://apnews.com/article/willow-oil-project-alaska-lawsuits-climate-aeb83859ad60f8ff4ef42eb2f6703a23}}) of recent news events, or choose another AP article they were interested in. 
Once the participants had read the article, they were asked to use the system to create a script.  
After a semi-structured interview about their script creation experience, the participant was instructed to move on to storyboarding.
Then they were asked to film the news reel in any way they felt most comfortable. 
Once the reel was complete, the participant was interviewed about their experience with ReelFramer, covering comments and suggestions on particular features and the overall creative process. 
This process was then repeated for a second article.

\subsection{Results}
\subsubsection{ReelFramer helps users effectively explore the design space for script creation.} 
All participants agreed that ReelFramer provided them with a flexible workflow to explore possible narrative framings, premises, and scripts. 
On average, it took participants 12.8 minutes ($SD$=4.0) to finalize a satisfactory script. 

Exploring narrative framings enabled users to find a fit for the tone of a news article.
All participants tried three framings in each of their script creation sessions.
P4 noted \textit{``The two articles I made reels for have very different styles; one is information-intensive, and the other is more story-driven. I like how the system provides three distinct framings for me to try, so I could see which fits best.''}  
Generating multiple premises and scripts helps users with divergent thinking.
P2 commented that within the comedic analogy framing, it was valuable to try multiple premises:  
\textit{``I like that it gives you three analogies to work with. 
I could pick the one that I feel most appropriate to continue with.''} 
By regenerating the scripts, P3 saw new possibilities for delivering the information, \textit{``I clicked the regenerate script button five times, each time a different script popped up. It was amazing; I definitely saw a bunch of options there.''} 

\subsubsection{Premises help users control generations and create coherent, informative scripts.}

All participants reported that ReelFramer greatly eased their creative process by offering meaningful scripts with an intermediate premise step.
Four participants mentioned that the intermediate premise step was useful. 
P5 noted, \textit{``It was really helpful to have all the key points listed. Thus, you know who your characters are, what your plot is, and what information you want to emphasize before you go writing the real dialogue.''} 
Participants all agreed that the provided premises were generally good.
Two participants commented that the information points extracted by ReelFramer were \textit{``very accurate''} (P4) and \textit{``capture the main takeaways I want to cover when reading the article''} (P1). 
While P5 said the extracted information points were not perfect, the suggested bullet points still offered them a reasonable starting point to break down the dense news print.

Three participants said ReelFramer gave appropriate character combinations. P4 said it helped \textit{``find the combination that is the crossfire in the article''}.
P2 found the character suggestions from the system \textit{``unexpected and funny''}---\textit{``The system suggested the oil company be a naïve newcomer (for expository dialog). 
That’s pretty funny because usually, we wouldn’t think of the oil company as naïve, but I like it and decide to roll with it.''}

Even if sometimes the premise generations were not perfect, participants appreciated that ReelFramer allowed them to take control and edit. 
P3 said, \textit{``Deciding the facts to include is where I am most nit-picky, and the system allows me to edit them directly, which is very convenient.''} 
P3 also liked how they can control the storyline and check the different script deliveries---\textit{``When I keep the premise and click `generate script,' their storylines were similar, but the delivery was different each time.'' }

\subsubsection{Script generation reduces time and effort.}
All participants agreed the generated scripts were generally coherent and informative.
Automatically generating scripts was a huge time saver.
\textit{``The scripting itself is definitely the most tedious and time-consuming part''} as P1 said, \textit{``The system did a pretty good job of suggesting draft scripts, I would say for the most part like 90\%.''}  
Participants commented that the generated scripts were generally coherent with \textit{``realistic dialogues''} (P1), \textit{``sophisticated exchanges and good comebacks''} (P2), and \textit{``digestible information explanations provided''} (P5). 
As a creator who had made news TikToks before, P2 said it was much easier and more fun to make news reels with ReelFramer than their previous experience---\textit{``Lots of the scripts generated by the system are so interesting, I would love to use it to create more news reels in the future.''}
However, the script generations were not perfect; for example, as P2 revealed, sometimes the generated drafts were not as funny as expected---even those scripts needed editing, automatic script generation provided big time-saving.

\subsubsection{In human-AI co-creation, AI can ideate and generate, but humans need to evaluate and edit at every step to ensure quality.}
All participants reported editing the premise and script. Although the AI suggestions were helpful, they had to be edited for consistency, coherency, and appropriate tone.

In terms of the premise editing, P2 changed the characters to make sure they aligned with the comedic plot, P3 replaced the suggested character with another that they felt more people knew, and P5 added references in the news article as characters to make the undertone of the script match that of the article. 
P1 slightly refined an information point to simplify the language---they used familiar concepts like ``reduce greenhouse gas emissions'' to replace the more technical terms.
P5 added figures and statistics to the information points because they believed it was important for the audience to know the scale of an issue.

Even with a good premise, all participants said scripts needed editing, to make them fit for human audiences. 
Three participants (P1-P3) reported that they adjusted the punchlines as sometimes they were \textit{``a little bit too cringy''} (P1) and \textit{``need a better clarification''} (P3). 
P1 adjusted the representation of information in the script to be less confusing, 
P2 edited the draft to make the dynamics fully flesh out the main characters, 
P5 normalized the tone of the scripts to make them more neutral as the original article, 
P3 and P4 did word-level edits to make the conversation a bit more colloquial. 

Evaluating AI outputs was a major role for the human to play in the co-creative process. 
Tools to reduce the effort of evaluation helped. 
Premises helped reduce the number of scripts to read and P1 and P5 commented that the highlight feature helped evaluate the news information in the script.

\subsubsection{Character boards help ideate visual elements}
Four of the five participants said the storyboarding features were not necessary for them. 
They had shooting experience before (three shot TikToks before and one was enrolled in a drama major) and were able to build a mental image of the shots, expressions, and actions just based on the script. 
However, they still found the character board as a useful foundation for how to visually depict the characters. 

For the character boards, all the participants agreed that the character descriptions were accurate and inspiring. 
For example, P1 commented \textit{``I like how the system gives different personalities to the two banks. For example, a young person in a West Coast style representing Silicon Valley Bank and a person in more traditional clothing representing Signature Bank.''} 
Even if users cannot exactly follow the prop and costume suggestions, they still took inspiration---\textit{``Although I don't have the same outfits right now, the character board gave me the idea that I should be wearing more formal attire [to represent a bank].''} (P1). 
All participants except P2 agreed that the background images were accurate. 
Most participants took the idea of the suggestions to find stock images, and one participant (P1) directly used the suggested images as the virtual background of their news reels.

\subsubsection{Overall impression and suggestions}
Overall, ReelFramer lowers the barrier for journalists to create news reels.
Four participants were first-time news reel creators; they said they would love to publish the reels they created and use ReelFramer in the future. 
One participant made news reels before and commented that ReelFramer greatly improved the efficiency compared to their previous manual way and they would love to incorporate ReelFramer into their future workflow. 
One of the four freshman reel creators attended a career fair a few days after the study. 
Several news outlets were looking for people with experience making news reels. 
Because of their ReelFramer experience, they applied for the position.

During the interview, participants also provided feedback on future improvements.
P1, P4, and P5 suggest that we better incorporate the original article into the system. 
To refine the information points to be included in the script, P1 and P5 went back and forth between the system and the article several times. 
This is tedious, and there are several ways to help users stay closer to the articles. 
For example, instead of summarizing information points, the system can extract lines from the article, and link back to it. 
\section{Expert Evaluation of News Reels}
We invited journalism experts and TikTok power users (actively engaged with TikTok for at least three hours daily) to evaluate the news reels that our user study participants created. 
This evaluation aims to understand whether the reels are accurate and informative (from journalism experts) and whether they are coherent, entertaining, and fit for TikTok (from TikTok power users). 

\subsection{Journalism Experts' Perspectives}

\begin{figure}[!t]
    \centering
    \includegraphics[width=0.43\textwidth]{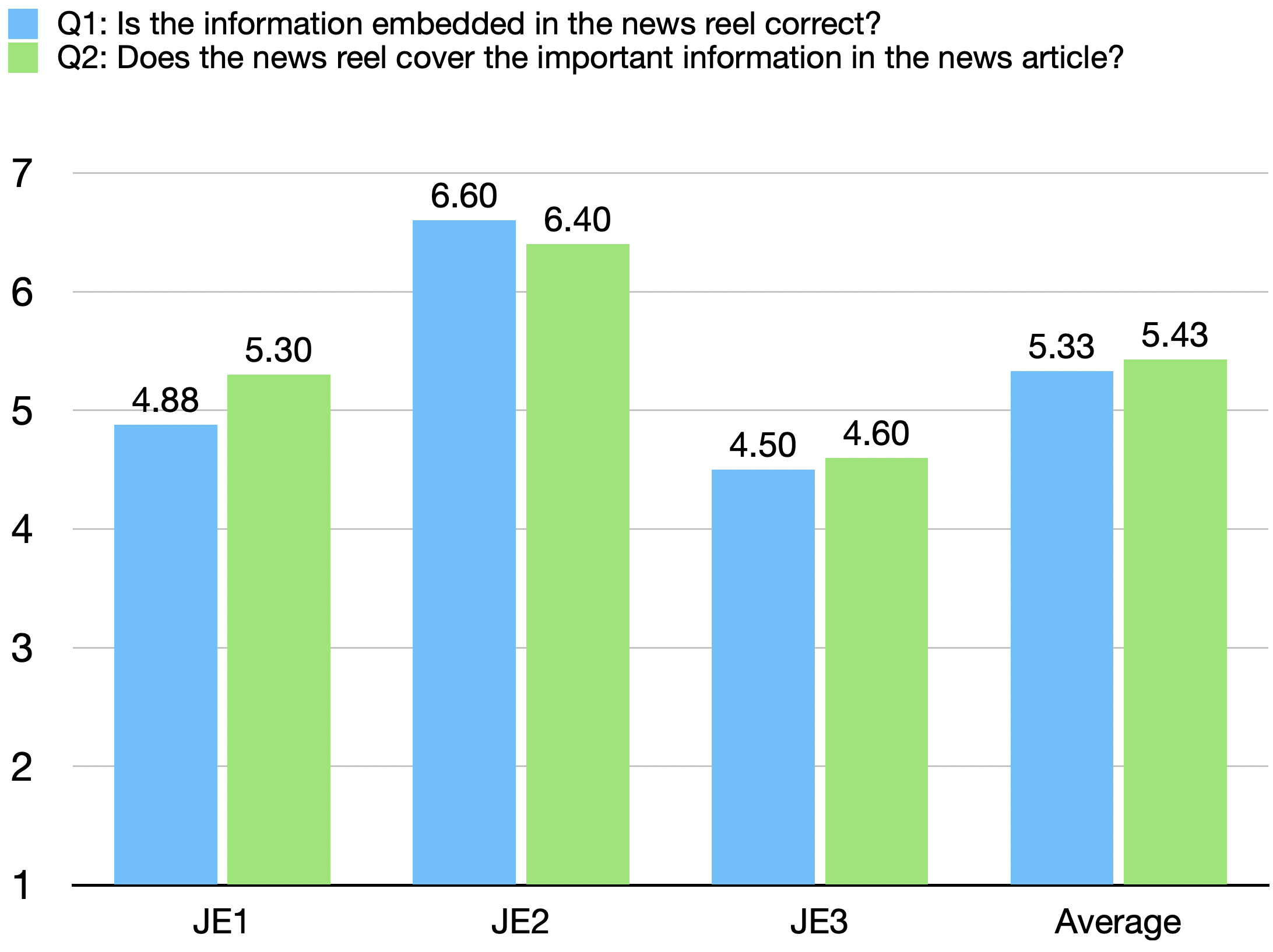}
    \caption{Average ratings provided by journalism experts for the accuracy and comprehensiveness of the news reels.}
    \label{fig:video_eval_experts}
    \Description{A bar chart showing how three journalism experts rate the final output videos through two major questions related to information accuracy. The two questions are Q1. Is the information embedded in the news reel correct? Q2. Does the news reel cover the important information in the news reel? For the specific scoring, JE1 gives an average of 4.88 for Q1 and 5.30 for Q2; JE2 gives an average of 6.60 for Q1 and 6.40 for Q2; JE3 gives an average of 4.50 for Q1 and 4.60 for Q2. Across the three experts, the final average ratings for Q1 are 5.33 and 5.43 for Q2.}
\end{figure}

We hired three journalism experts (one female, two male; average age 26.7, different from previous studies' participants) to evaluate the informative aspect of the reels. 
They were all experienced journalists who produced award-winning work.
Additionally, they were all avid TikTok users.
We asked them to first carefully read the news articles and then review the reels based on two questions:
\begin{itemize}
    \item Q1: Is the information embedded in the news reel correct?
    \item  Q2: Does the news reel cover the important information in the news article? 
\end{itemize}
The journalism experts provided ratings on a 7-point scale (see Figure \ref{fig:video_eval_experts}) with justifications.

For Q1, the average score was 5.33 ($SD$=1.12), indicating that all reels performed strongly in terms of correctness. 
All experts agreed that most reels accurately represent information by successfully identifying the characters and their relationships (JE1, JE2, JE3), presenting numbers (JE1, JE2), and main arguments of the news (JE1, JE2, JE3). 
However, there were some aspects that were not entirely accurate. 
The discrepancy was mostly attributed to the artistic decisions made by creators. 
For example, a creator decided to portray an oil company as a person drilling into the Alaskan North Slope. 
Due to this, JE1 ranked this video a 4/7 on correctness, as this implied that the company has started operationalizing the oil extractive aspects when that is not the case. 
Lower-rated videos also stemmed from representing a group as a single character or portraying a well-informed stakeholder as naive. 

For Q2, the average score was 5.43 ($SD$=0.91), indicating that the reels were reasonably informative. 
All experts agreed that the reels hit almost all the important information points by capturing different viewpoints of the debate (JE3), presenting all major players and their roles (JE1, JE2), and bringing up related historical contexts (JE2).
However, these videos usually lack a bigger picture of the events. 
Experts shared that some reels only provide a superficial discussion of the event itself without presenting the underlying need and larger implications. 
Thus, JE3 rated a comedic analogy reel discussing the Willow Project conflicts in the context of fitness training, with a 3/7. 
They expressed concerns that the audience without prior knowledge about the news might not get the joke. 

\subsection{TikTok Power Users' Perspectives}
\begin{figure}[!t]
    \centering
    \includegraphics[width=0.43\textwidth]{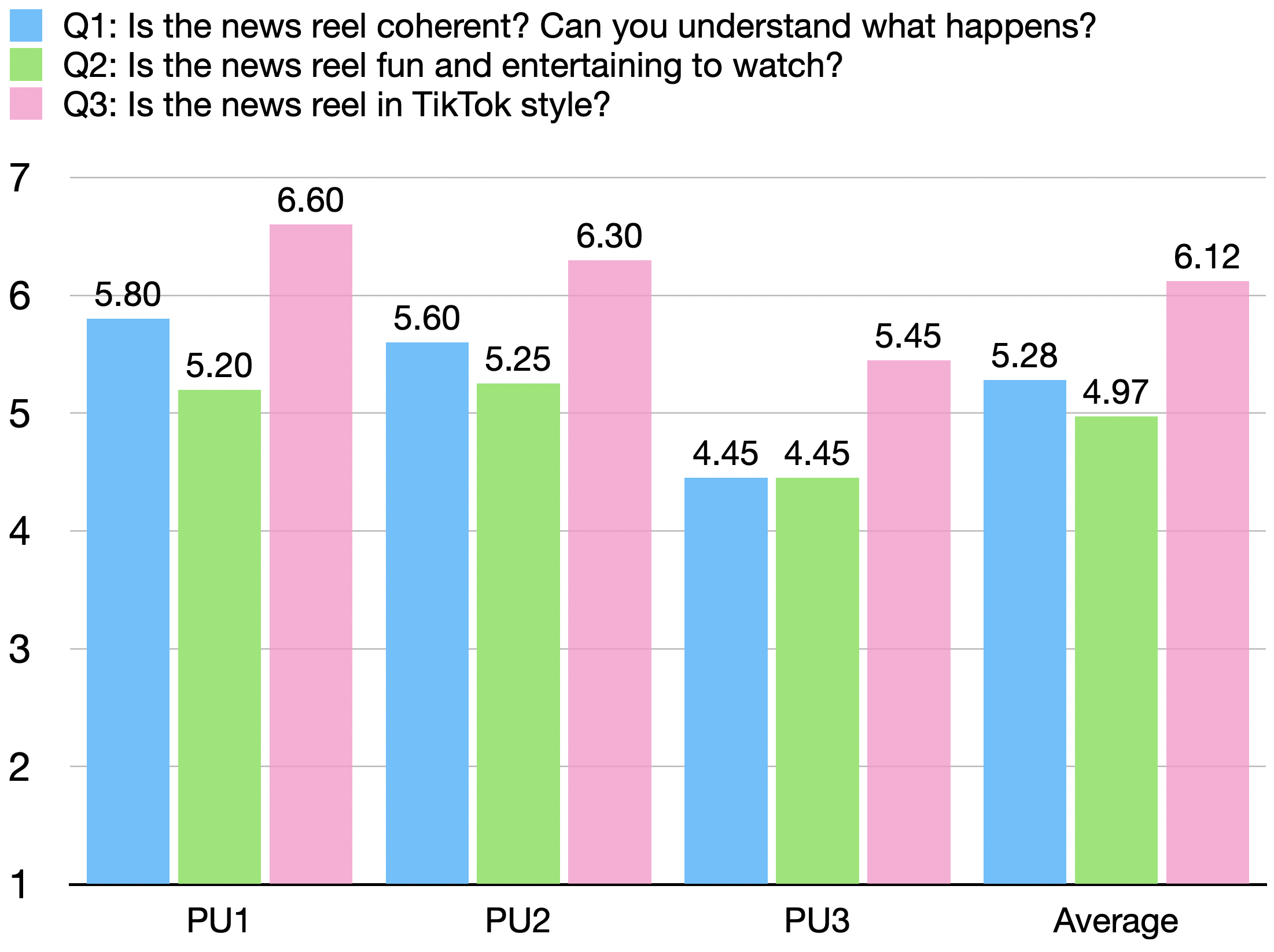}
    \caption{Average ratings provided by TikTok power users for the entertainment and style aspect of the news reels.}
    \label{fig:video_eval_tiktokusers}
    \Description{A bar chart showing how three TikTok power users rate the final output videos through three major entertainment-related questions. The three questions are Q1. Is the news reel coherent – can you understand what happens? Q2. Is the news reel fun and entertaining to watch? Q3. Is the news reel in TikTok style? For the specific scoring, PU1 gives an average of 5.80 for Q1, 5.20 for Q2, and 6.60 for Q3; PU2 gives an average of 5.60 for Q1, 5.25 for Q2, and 6.30 for Q3; PU3 gives an average of 4.45 for Q1, 4.45 for Q2, and 5.45 for Q3. Across the three users, the final average ratings for Q1 are 5.28, 4.97 for Q2, and 6.12 for Q3.}
\end{figure}

We hired three TikTok power users (one female, two male; average age 21.3, different from previous studies' participants) to evaluate the entertainment value of the reels. 
They were within the target audience of platforms like TikTok and familiar with how TikTok delivers news content. 
To emulate the experience of discovering a TikTok on the feed, we asked them to directly watch the reels without reading the original articles. 
We asked them three questions: 
\begin{itemize}
    \item Q1: Is the news reel coherent? Can you understand what happens?
    \item Q2: Is the news reel fun and entertaining to watch? 
    \item Q3: Is the news reel in TikTok style?
\end{itemize}
The TikTok power users provided ratings on a 7-point scale (see Figure~\ref{fig:video_eval_tiktokusers}) with justifications. 

For Q1, the average score was 5.28 ($SD$=0.59), showing that most reels were coherent and clear for the audience to easily follow. 
PU1 and PU3 shared that they could quickly identify the relationship between the characters and understand their standpoints with the aid of different body movements, props, and visual appearances. 
However, while PU2 shared that the color-coded textual description contributed to the coherence of the video, too many textual and visual elements on one screen can make the content difficult to absorb, resulting in lower ratings for two reels (3/7 and 4/7). 

For Q2, the average score was 4.97 ($SD$=0.37), showing that the audience found many reels fairly entertaining to watch. 
They shared that the clear and dramatic delivery of the reels contributes to enjoyment and fun, like the tone, voice, acting, language, and punchline (PU1, PU2).  
Also, PU2 shared that the funny background images and overlaying GIFs help with the storytelling. 
The drawbacks mentioned are some reels are too fact-heavy by stating many numbers (PU1) or too long (PU3).
Enjoyment also seemed to be based on the acting skill of the creator, with more energetic performances favored by the audience.
PU3 noted that even a mediocre script could be entertaining if the creator has ``great charisma.''

For Q3, the average score was 6.12 ($SD$=0.49). 
All users expressed that most reels captured the essence of TikTok reels, and would fit seamlessly into their feeds.
PU1 shared that the use of effects like text captions and the TikTok green-screen effect underlined key, lo-fi characteristics of TikTok content. 
PU2 even noted that the reels on the Willow Project were ``almost exactly like the videos I've seen on the Willow Project on TikTok.''
Overall, the reels effectively captured the TikTok style, while maintaining high coherence and moderate entertainment value. 
\section{Discussion}
\subsection{Narrative Framing for Content Retargeting}
ReelFramer provides narrative framing as a structure to support the translation from print articles to news reels that require both information and entertainment value. 
Our studies show that narrative framing introduces necessary diversity for the translation of different articles, and establishing narrative details of the framing helps generate scripts that are more relevant and coherent.

Narrative framing is a storytelling concept that guides how a story is structured and presented to shape the audience's perception and understanding~\cite{schmid2021narrative}.
ReelFramer formalizes the application of narrative framing in AI within journalism, which involves retargeting news content to appeal to new audiences.
Different news articles call for different framings.
ReelFramer allows journalists to explore framings that span the infotainment spectrum in a structured way.
Rather than asking users to decide on the framing at the very beginning~\cite{mirowski2022cowriting}, our method provides a more flexible way for users to explore the creative space.
To ensure the narrative framing can be successfully executed, the essential details need to be established before the final writing is created. 
This was also explored in recent co-writing systems~\cite{mirowski2022cowriting, VISAR}.
However, unlike in fictional writing, news reels must adhere to the facts in the article. 
Thus, besides traditional narrative elements~\cite{halliwell1998aristotle}---characters, plot, and setting, the narrative details in ReelFramer also include information points. 
This is to help users identify and present the essential fundamentals accurately and faithfully in reels, just as in their articles.
Establishing narrative details is essential for content retargeting tasks to ensure that the original information is properly translated.

We believe narrative framing can be adapted to other content retargeting tasks. 
For example, to create reels that introduce science, the three narrative framings can still apply: one can imagine a reenactment between an atom nucleus and electrons following around it, explaining the atom structure.
The narrative details, in this case, need some modification.
While characters, plot, and setting still fit, the core message to deliver is different. 
The foundational information in science reels might focus on scientific mechanisms instead of covering news information points as seen in news reels.

\subsection{Roles of Human and AI in the Co-Creation Process}
Based on this work, we believe that humans and AI have complementary strengths in the creative process.
Creativity entails both divergence and convergence~\cite{guilford1967nature}. 
In the divergent phase, the goal is to generate many ideas.
Humans often fail to ideate broadly because of the mental propensity to fixate on a limited number of ideas~\cite{jansson1991design}.
However, generative AI can produce a great number of ideas and help explore the design space. 
In the convergent phase, the goal is to evaluate, refine, and synthesize the generated ideas.
Currently, AI models cannot accurately evaluate the quality of their outputs; humans are adept at making heuristic judgments.
Thus, it is important to support a co-creation process that leverages the combination of human and machine intelligence. 

When humans and AI collaborate on creative tasks, it is essential for humans to maintain control.
For example, during the premise generation step, AI can quickly generate lots of comedic analogies.
This provides users with multiple options to start with, reducing the cognitive demands of brainstorming on their own.
However, although users generally appreciate the suggestions from AI, the generated analogies sometimes go too far and/or are inappropriate.  
Directly using these analogies for news infotainment can be misleading. 
Human input is thus necessary to keep an appropriate tone of content, which is tricky to complete by AI alone. 

When considering a metaphor for the role of AI in co-creation, AI in ReelFramer is similar to that of an ``apprentice/assistant'' within the apprentice framework~\cite{apprentice}.
This implies that AI can generate a reduced set of prototypes that are correct examples of the desired kind and exhibit creative properties, such as being surprising, as noted in feedback from our user study participants. 
However, AI is not a flawless ``apprentice/assistant''; 
it cannot always produce accurate artifacts or be surprising, let alone attain the level of a ``master'' capable of performing all important tasks and delivering a complete, finished product.
In ReelFramer, AI always takes the first turn to explore the creative artifact~\cite{guzdial2019interaction} while people can concentrate on assessing and improving the quality of the work.
This is called ``task-divided creativity'' as AI here is an incomplete agent that cannot independently complete all the creative tasks such as evaluation~\cite{kantosalo2016modes}.

\subsection{Limitations}
ReelFramer was co-designed with journalists to ensure it aligns with journalistic practices and values. 
This collaborative process included two experienced journalism instructors. 
Expanding the co-design to include the input of young journalists, however, might enrich the human-AI interaction paradigm by incorporating their distinctive perspectives. 
As many young journalists are active social media users, they may have specific needs to consider, such as a preference for memes, as indicated by the user study. 
Additionally, the user study featured a relatively limited number of participants. 
By involving more target users from various journalism areas (e.g., fashion and technology), we can gain a broader understanding of the diverse requirements across the profession.

ReelFramer has limitations regarding the provided structures.
While structures can introduce similarities, we did not observe a significant amount of similarity among the reels produced in our user study. 
This is because participants were able to adapt the provided frameworks in their own unique ways. 
However, it is important to note that all reels created in the study were roleplays.
This does limit the stories journalists can tell; for example, roleplays cannot be applied to very serious news issues.
By analyzing news-related fields, we can broaden the range of topics to cover.

The underlying technologies of ReelFramer have limitations in terms of accuracy.
GPT-4 has improved in accuracy compared to its predecessors, but it can still hallucinate~\cite{gpt4}. 
Since all our user study participants have journalism training, they naturally fact-check all the generations.
Thus, they corrected the false or misleading statements early and made sure the final artifact accurately portrayed the facts.
However, if users lack experience and training in fact-checking, there is a risk that the reels could be misleading.
To mitigate this, we could provide fact-checking tools in the system and nudge users towards using them. A simple first step would be to encourage users to compare their script with relevant excerpts from the article, and flag lines that have significant differences.

Additionally, biases are present in generative AI models. 
GPT-4 might exhibit a preference for certain genders or races when designing characters~\cite{kotek2023gender}; 
For instance, bankers tend to be depicted as male, while bank representatives often appear as female. 
Similarly, DALL-E 2 could represent characters in ways that reinforce stereotypes~\cite{cho2023dall}. 
For example, when asked to portray an Alaska native, DALL-E 2 might generate an image featuring stereotypical tribal attire with a feathered headdress, which may reinforce inaccurate cultural perceptions.
To additionally support humans in recognizing and correcting the biases, we can implement nudges that prompt users to consider whether there is any bias in the generated content, such as contextual warnings or reminders to check for diverse perspectives.

\subsection{Future Work}
\label{future_work}
In the future, we would like to extend ReelFramer to integrate more social media trends. 
TikTok is famous for its trending songs and visual gags~\cite{vizcaino2022music}. 
On TikTok, trending songs often serve as a backdrop, setting the emotional tone and rhythm for the reels.
For ReelFramer, which currently focuses on semantic and visual elements, incorporating the use of trending songs is a natural next step.
Visual gags involving props, filters, and stickers also play a critical role in successful reels by enhancing engagement and storytelling. 
While ReelFramer focuses on physical props for character portrayal, virtual options are more feasible and potentially more interesting.
Both trending songs and visual gags can be optional elements added to the premise to explore.

ReelFramer enables users to create their first news reels; however, the long-term value it offers remains unclear.
One might wonder if the system's usefulness is limited to initial creations, or if it can deliver lasting benefits for ongoing use. 
Further questions to be considered include whether the system can adapt to the evolving needs of the users and whether it allows for customization to enable users to express their own voices.
In the future, we plan to conduct a month-long co-design study in which participants will actively use the system. 
The study can provide insights into how ReelFramer can be tailored to better serve users’ unique expressive needs. 
In collaboration with instructors, we also hope to introduce the system to journalism classes, teaching more young journalists to create their first news reels.
\section{Conclusion}
We present ReelFramer, a human-AI co-creation system that supports journalists in retargeting news articles into reels. 
ReelFramer consists of two stages to streamline the ideating and prototyping process of the news reel creation: scriptwriting and storyboarding. 
The system offers three narrative framings to balance the reel's informational and entertainment values while supporting the exploration of their narrative details.
It also facilitates the visual exploration of character design and storyboarding.
Our study shows that ReelFramer helps users effectively explore the design space and lowers the barriers to making their first news reels.
We discuss what roles humans and AI should take in the co-creative process.

\begin{acks}
This work is supported by NSF-IIS-2129020, NSF-IIS-2128906 and NSF-IIS-2129047.
\end{acks}

\bibliographystyle{ACM-Reference-Format}
\bibliography{sample-base}

\appendix
\clearpage
\section{ReelFramer prompts}
\label{appendix:prompts}
The news article is attached to the beginning of each prompt from Section \ref{appendix:news} to \ref{appendix:scripts}.
\subsection{Extract news information}
\label{appendix:news}
\begin{itemize}
    \item \textit{news setting}: ``Where did this news event take place?'' 
    \item \textit{news characters} and their main activities: ``List names of the five main stakeholders in this news event and what they mainly did.''
    \item \textit{news plot summary}: ``What happened in the news event?'' 
    \item \textit{news information points}: ``What are the three most important things in this news story?'' 
     \item \textit{news plot elements}: ``What are the four main plot points of the news story?''
\end{itemize}

\subsection{Get script premises}
\subsubsection{Expository dialog}
\begin{itemize}
    \item \textit{script plot}: No prompt needed, always ``a person (expert) explaining the information to another (naive newcomer)''
    \item \textit{script characters}: ``The main characters of the news event are \textit{news characters}, which two are expert and naive newcomer?'' 
    \item \textit{script setting}: ``To make a short video of this news event on social media, where might the location be?''
    \item \textit{information points}: No prompt needed, directly use \textit{news information points} extracted in C.1
\end{itemize}

\subsubsection{Reenactment}
\begin{itemize}
    \item \textit{script plot}: No prompt needed, ``Key stakeholders acting out what happens---\textit{news plot summary}'' (use \textit{news plot summary} extracted in C.1)
    \item \textit{script characters}: ``The main characters of the news event are \textit{news characters}, based on what happened---\textit{news plot summary}, which two are the most dominant characters in the news event?''
    \item \textit{script setting}: ``To make a short video of this news event on social media, where might the location be?''
    \item \textit{information points}: No prompt needed, directly use \textit{news information points} extracted in C.1
\end{itemize}

\subsubsection{Comedic analogy}
\begin{itemize}
    \item \textit{script plot}: ``List three unique comedic analogies for the situation in the following story: \textit{news plot summary}. Incorporate the following characters only: \textit{script characters}''
    \item \textit{script characters}: ``The main characters of the news event are \textit{news characters}, based on what happened---\textit{news plot summary}, which two are the most dominant characters in the news event?''
    \item \textit{script setting}: ``To act out this analogous premise \textit{script plot}, where is the location?''
    \item \textit{information points}: No prompt needed, directly use \textit{news plot points} extracted in C.1
\end{itemize}
Note that to parse the premise automatically, we also attach the formatting instructions to each prompt above. The instruction defines the output format, such as ``Format should be Setting: [Setting].''

\subsection{Get Scripts}
\label{appendix:scripts}
\begin{itemize}
\item \textit{script}: ``Write a script for a comedy skit about \textit{script plot}. Cover the following information: \textit{script information points}. The character should be exactly \textit{script characters}. It should be set in \textit{script setting}.
It should be entertaining. The dialogue should be colloquial and engaging. The dialogue should be 10 to 12 lines long. Each line of dialogue should be short---less than 20 words. End it with a punchline.''
\end{itemize}
Note that to parse the script into a readable format in the system, we also attach the following instructions to facilitate the parsing: ``Put parenthetical between parentheses. Put non-dialog parts between square brackets. Capitalize the character names and put a colon after. Separate each dialog line with double new line characters.''

\subsection{Character board}

\subsubsection{Large language model prompts}
\begin{itemize}
\item \textit{character descriptions}: ``You are a costume designer for a film. Considering this plot \textit{script plot} and this screenplay \textit{script}. Describe a single person who could represent each of the characters or groups and map them to one of the characters in this list \textit{script characters} in a numbered list. [character] is a [man/woman/person] who wears [item].''

\item \textit{character props}: ``Considering these characters \textit{character descriptions}. Write a list of two clothing items and two household props that an actor could wear to play each character.''

\item \textit{visual setting}: ``You are an art director for a film. Considering this screenplay \textit{script}, choose a setting.'' 

\item \textit{background description} ``Considering this setting \textit{visual setting} and this premise \textit{script premise}, for each of the following characters \textit{character descriptions}, list where they would be in this setting, and what the area behind them would look like.''

\item  \textit{background image prompt} ``Write a prompt for a generative art program to create an image of this background area based in \textit{visual setting}: \textit{background description}.''
\end{itemize}

\subsubsection{Text-to-image model prompts}
\begin{itemize}
\item \textit{character images}: ``A waist-up portrait of \textit{character descriptions}, with \textit{character props}, in the style of black-and-white vector line art.''

\item \textit{background images}: ``\textit{background image prompt}, in the style of a digital painting background.'' 
\end{itemize}

\subsection{Storyboard}
\subsubsection{Large language model prompts}
\begin{itemize}
    \item \textit{expression}, \textit{gesture}, \textit{action}: ``\textit{script line} To act out this script line:  give one key phrase for expression, gesture, and action.'' 
\end{itemize}

\subsubsection{Text-to-image model prompts}
\begin{itemize}
\item \textit{storyboard images}: ``\textit{character descriptions}, looking \textit{expression}, \textit{gesture}, \textit{action}, in the style of black-and-white vector line art'' and generate images for each shot.
\end{itemize}

\end{document}